\documentclass[aps,showpacs,pre,superscriptaddress,onecolumn,nofootinbib]{revtex4}
\usepackage{amsfonts}
\usepackage{amsmath}
\usepackage{mathrsfs}
\usepackage{amssymb}
\usepackage{wasysym}
\usepackage{subfigure}
\usepackage{graphicx}
\usepackage{CJK}
\usepackage{float}
\usepackage[toc,page,title,titletoc,header]{appendix}
\usepackage{color}

\begin{document}

\title{Ginzburg-Landau effective action approach to disordered Bose-Hubbard Model}
\author{Botao Wang}
\affiliation{Department of Physics, Shanghai University, Shanghai 200444, P.R. China}
\author{Jun Zhang}
\affiliation{Department of Physics, Shanghai University, Shanghai 200444, P.R. China}
\author{Ying Jiang}
\thanks{Corresponding author}
\email{yjiang@shu.edu.cn}
\affiliation{Department of Physics, Shanghai University, Shanghai 200444, P.R. China}
\affiliation{Shanghai Key Laboratory of High Temperature Superconductors, Shanghai 200444, P.R. China}

\begin{abstract}
We study the phase transition from Mott insulator (MI) to Bose glass (BG) of a disordered Bose-Hubbard model within the framework of Ginzburg-Landau effective action approach.
By treating MI as unperturbed ground state and performing a systematic expansion with respect to tunneling matrix element, we extend such a field-theoretic method into the disordered lattice Bose systems.
To the lowest order, a second order phase transition is confirmed to happen here and the corresponding phase boundary equation coincides with the previous mean-field approximation result.
Keeping all the terms second order in hopping parameter, we obtain the beyond mean-field results of MI-BG phase boundary of 2D and 3D disordered Bose-Hubbard models.
Our analytic predictions are in agreement with recent semianalytic results.

\end{abstract}
\maketitle

\section{Introduction}

The interplay between disorder and interaction has been an important yet challenging topic in real condensed matter systems \cite{1958Anderson,1978Mott,1989Fisher}.
Since the pioneer work of observing the superfluid (SF) to Mott insulator (MI) phase transition in optical lattice systems \cite{2002MGreiner}, where the parameters possess a broad tunability, it becomes a natural extension to realize disordered systems by using ultracold atoms in optical lattices \cite{2008IBloch,2008LFallani,2012MLewenstein}.
Nowadays, random potential in optical lattice can be created in many different ways, e.g. using speckle patterns \cite{1999DBoiron-disorder,2005JELye,2010LSPalenciaSpeckle}, multi-chromatic incommensurate optical lattices \cite{2007LFallani}, atomic mixtures \cite{2005UGavish-mixtureDisorder,2011BGadway} and so forth.
The introducing of disorder can give rise to the appearance of Bose glass (BG) phase, an insulating phase characterized by a finite compressibility, no gap, but an infinite superfluid susceptibility \cite{1989Fisher}.
An ongoing effort has been devoted to the disordered lattice systems \cite{2008IBloch,2008LFallani,2012MLewenstein,2011BGadway,2010MPasienski,2014CDErrico}, and until recently the phase transition of SF-BG can be detected in three-dimensional (3D) optical lattices \cite{2016CMeldgin-BG}.
Although an insulating regime at strong interaction and strong disorder has also been detected \cite{2007LFallani,2011BGadway,2010MPasienski}, so far, it remains a big challenge to distinguish the BG phase from the MI phase, especially in 2D and 3D optical lattices \cite{2014CDErrico}.

As the first successful attempt at the description for disordered strongly-interacting bosonic systems, the disordered Bose-Hubbard model (BHM) has been theoretically introduced by M. Fisher \emph{et al.} in 1989 \cite{1989Fisher}.
Since then the phase diagram of disordered BHM has been studied with a wide range of methods \cite{1991Krauth-direct2D,1997Kisker,2001Lee,1991Scalettar-MI1D,2004Nprokof,2009VGurarie3DQMC,2011Lin,2011SGSoyler,2015CZhang-QMC,1996Svistunov-RG,2011Kruger,2013Hegg,2010Carrasquilla-Probability,2012CHLin,1999Rapsch,2009Deng,1996Freeriscks,2013AENiederle,2009Bissbort-SMF,2006Krutitsky,2007Buonsante-MF,2011Pisarski-MF}, including quantum Monte Carlo (QMC) simulations \cite{1991Krauth-direct2D,1997Kisker,2001Lee,1991Scalettar-MI1D,2004Nprokof,2009VGurarie3DQMC,2011Lin,2011SGSoyler,2015CZhang-QMC}, density-matrix renormalization group (DMRG) techniques \cite{1999Rapsch,2009Deng}, strong coupling expansion \cite{1996Freeriscks}, local mean-field cluster analysis \cite{2013AENiederle} and many other mean field theories \cite{2009Bissbort-SMF,2006Krutitsky,2007Buonsante-MF,2011Pisarski-MF}.
On the numerical side, QMC has been able to determine the SF-BG phase boundary in strong interaction regime \cite{2009VGurarie3DQMC,2011SGSoyler,2015CZhang-QMC}. Nevertheless, since the occurrence of the BG phase is caused by the arbitrarily large but exponentially rare SF puddles within the MI background \cite{1989Fisher}, the compressibility of BG is so small that it is impossible to directly distinguish between the MI and BG phases in finite-size simulations \cite{2009VGurarie3DQMC,2011SGSoyler,2015CZhang-QMC}.
Analytically, most studies have been limited to a mean-field level \cite{2013AENiederle,2009Bissbort-SMF,2006Krutitsky,2007Buonsante-MF,2011Pisarski-MF}, while the equilibrium properties of disordered BHM has been seldom addressed beyond mean-field theory so far.

Based on a field-theoretical consideration, the so-called Ginzburg-Landau effective action approach has been worked out \cite{2009Santos,2009Bradlyn} and widely applied to investigate the phase transition in clean lattice Bose systems \cite{2012ZLin,2012CNietner-JCM,2013MMobarak-spin1,2013TWang,2015XFZhang}.
Besides the advantages of yielding analytical results at arbitrary dimension and lobe number, it can also lead to arbitrarily high accuracy as higher hopping orders are evaluated, which turns out to be in excellent agreement with QMC simulations \cite{2009AEckardt,2009NTeichmann-process-chain,2016Wei}.
Considering its big success in determining the phase diagram of clean BHM \cite{2009Santos,2009Bradlyn,2012ZLin,2012CNietner-JCM,2013MMobarak-spin1,2013TWang,2015XFZhang,2009AEckardt,2009NTeichmann-process-chain,2016Wei}, in this work we extend this method to the disordered BHM. To this end the beyond mean-field analytic results of MI-BG phase boundary in 2D and 3D are obtained.

We organize this paper as follows.
In Section II, we introduce the disordered Bose-Hubbard Hamiltonian and give a brief discussion at atomic limit. We develop the Ginzburg-Landau theory and give the effective action of disordered BHM with the help of cumulant expansion in Section III. In Section IV, we present our zero temperature phase diagrams of MI-BG in 2D and 3D, and the comparisons with the results coming from other methods are also conducted. Finally, we give a summary in section V.

\section{The Model}

The analysis of disordered ultradcold Bose gases trapped in an optical lattice is usually based on the so-called Bose-Hubbard Hamiltonian with on-site disorder \cite{1989Fisher,1998DJaksch}, which takes the form of $\hat{H}=\hat{H}_{0}+\hat{H}_{1}$ with

\begin{align}
\hat{H}_{0}=& {\displaystyle \sum_{i}\left[\frac{U}{2}{\displaystyle \hat{n}_{i}\left(\hat{n}_{i}-1\right)}-\mu_{i}\hat{n}_{i}\right]},
\\
\hat{H}_{1}=& -{\displaystyle \sum_{\left\langle i,j\right\rangle }t_{ij}\hat{a}_{i}^{\dagger}\hat{a}_{j}},
\label{H}
\end{align}
where $\hat{a}_{i}$ ($\hat{a}^{\dagger}_{i}$) is the bosonic annihilation (creation) operator, and $\hat{n}_{i}=\hat{a}_{i}^{\dagger}\hat{a}_{i}$ is the particle number operator on the lattice site $i$.
$U$ parameterizes the on-site repulsive interaction energy between the two atoms at a given site. $t_{ij}$ is the hopping parameter, which is symmetric in $i$ and $j$, satisfying $t_{ij}=t_{ji}=t$ and $t_{ii}=0$. Here we only consider the tunneling between the nearest neighbors, which is denoted by $\left\langle i,j\right\rangle$.
The on-site energy $\mu_{i}$ satisfies $\mu_{i}=\mu+\delta\mu_{i}$, where $\mu$ is the chemical potential. $\delta\mu_{i}$  describes the random potential, which is normally assumed to be spatially uncorrelated and uniformly distributed between $[-\Delta,\Delta]$, i.e.
\begin{align}
  \overline{\delta\mu_{i}}=0,~~\overline{\delta\mu_{i}\delta\mu_{i^{\prime}}}=\frac{\Delta^{2}}{3}\delta_{i,i^{\prime}}.
  \label{P-disorder}
\end{align}
Here $\overline{\cdots}=\frac{1}{2\Delta}\int_{-\Delta}^{\Delta}\cdots d\delta\mu_{i}$ stands for the disorder ensemble average and $\Delta$ represents the disorder strength.

Let us begin with a brief discussion in atomic limit ($t_{ij}\rightarrow0$), where the full Hamiltonian $\hat{H}$ reduces to the local one $\hat{H}_{0}$ with the energy eigenvalues
\begin{equation}
E_{\{n_{i}\}}={\displaystyle \sum_{i}}f_{i}(n_{i}),~~f_{i}(n_{i})\equiv\frac{U}{2}{\displaystyle n_{i}\left(n_{i}-1\right)}-\mu_{i}n_{i}.
\label{fn}
\end{equation}
It is simple to verify that under the condition $\Delta\leq U/2$, for any $\mu$ in the interval between $nU-\Delta$ and $\left(n-1\right)U+\Delta$, each site $i$ should contains exactly $n$ bosons to minimize the on-site energy, i.e. $f_{i}\left(n\right)\leq f_{i}\left(n-1\right),f_{i}\left(n\right)\leq f_{i}\left(n+1\right)$.
In this case, the system will therefore stabilize at incompressible MI states.
Note that the effect of disorder is to produce gaps of width $2\Delta$ between different MI lobes, where a new insulating BG state appears \cite{1989Fisher}.

While tuning on the hopping parameter, bosons can gain kinetic energy from hopping between nearest-neighbor sites. Therefore the localized states will be softened by the quantum fluctuations. In fact, with the increase of $t/U$, the complicated competition among hopping, on-site interaction and random potential will give rise to other different phases, i.e. SF phase and BG phase \cite{1989Fisher}.
In the present work we will try to determine the phase boundary between MI and BG phase within the framework of Ginzburg-Landau theory.
In a general field-theoretical way \cite{2009Bradlyn,2012CNietner-JCM,2001Kleinert,2002Justin}, the source terms $j_{i}\left(\tau\right), j^{*}_{i}\left(\tau\right)$ are artificially introduced to break global symmetries,
\begin{align}
\hat{H}_{1}\rightarrow\hat{H}_{1}^{\prime}\left(\tau\right)=-{\displaystyle \sum_{i,j}t_{ij}\hat{a}_{i}^{\dagger}\hat{a}_{j}}+\sum_{i}\left[j_{i}\left(\tau\right)\hat{a}_{i}^{\dagger}+j_{i}^{*}\left(\tau\right)\hat{a}_{i}\right],
\label{H1-j}
\end{align}
where $j_{i}\left(\tau\right), j^{*}_{i}\left(\tau\right)$ are assumed to be dependent on both space and time, but independent of individual disorder distribution.
Since we will take the limit of vanishing currents ($j_{i}\left(\tau\right), j^{*}_{i}\left(\tau\right)\rightarrow0$) in the end and the hopping amplitudes $t_{ij}$ is much smaller than the on-site interaction $U$ in the strongly interacting regime, we can treat both hopping and source terms in Eq. (\ref{H1-j}) as small quantities, indicating that $\hat{H}_{1}^{\prime}\left(\tau\right)$ will be treated as a perturbation in the following.

\section{Ginzburg-Landau effective action approach}

To first order in the tunneling, a Ginzburg-Landau expansion of the effective action for a general bosonic lattice Hamiltonian has been derived \cite{2009Bradlyn}. Since the primary advantage of such an effective-action theory lies in its extensibility, following a similar pace we apply this method in our disorder case, and to this end a higher-order correction in tunneling will be included.
As for the new Hamiltonian $\hat{H}\left[j,j^{*}\right]=\hat{H}_{0}+\hat{H}_{1}^{\prime}\left(\tau\right)$, the partition function is generally defined as
\begin{equation}
\mathcal{Z}\left[j,j^{*}\right]\equiv\mathrm{Tr}\left\{ \hat{T}e^{-\int_{0}^{\beta}d\tau\hat{H}\left[j,j^{*}\right]}\right\},
\label{Z-definition}
\end{equation}
which can be reformulated in Dirac picture as
\begin{equation}
\mathcal{Z}\left[j,j^{*}\right]=\mathrm{Tr}\left\{e^{-\beta\hat{H}_0}\hat{U}_\mathrm{D}(\beta,0)\right\}.
\label{Z-U}
\end{equation}
Here, $\hat{U}_\mathrm{D}(\beta,0)$  is the time-evolution operator in imaginary-time Dirac picture
\begin{align}
\hat{U}_{D}\left(\beta,0\right)=\hat{T}\exp\left\{ -\int_{0}^{\beta}d\tau_{n}\hat{H}_{1D}^{\prime}\left(\tau\right)\right\}
 ,\label{UD-n}
\end{align}
where $\tau=it$ is the imaginary time, $\hbar$ is set to be 1 and $\hat{T}$ is the imaginary-time ordering operator. All the operators which depend on imaginary time are taken in Dirac picture, i.e. $\hat{a}_{i}^{\dagger}\left(\tau\right)=e^{\tau\hat{H}_{0}}\hat{a}_{i}^{\dagger}e^{-\tau\hat{H}_{0}},~\hat{a}_{i}\left(\tau\right)=e^{\tau\hat{H}_{0}}\hat{a}_{i}e^{-\tau\hat{H}_{0}}$.
Introducing $
\left\langle \bullet\right\rangle {}_{0}=\frac{1}{\mathcal{Z}^{(0)}}\mathrm{Tr}\left\{ e^{-\beta\hat{H}_{0}}\bullet\right\} $
for the thermal average with respect to the unperturbed Hamiltonian $\hat{H}_0$ and substituting Eq. (\ref{UD-n}) into Eq.~(\ref{Z-U}), we expand the partition function as
\begin{align}
\mathcal{Z}\left[j,j^{*}\right] & =\mathcal{Z}^{(0)}+\sum_{n=1}^{\infty}{\mathcal{Z}^{(n)}\left[j,j^{*}\right]},
\nonumber \\
\mathcal{Z}^{(n)}\left[j,j^{*}\right] & =\mathcal{Z}^{(0)}\frac{(-1)^{n}}{n!}\int_{0}^{\beta}d\tau_{1}\int_{0}^{\beta}d\tau_{2}\dots\int_{0}^{\beta}d\tau_{n}\left<\hat{T}\left[\hat{H}_{1\mathrm{D}}^{\prime}(\tau_{1})\hat{H}_{1\mathrm{D}}^{\prime}(\tau_{2})\dots\hat{H}_{1\mathrm{D}}^{\prime}(\tau_{n})\right]\right>_{0},
\label{Z_n}
\end{align}
where
\begin{align}
\mathcal{Z}^{(0)} =\mathrm{Tr}\{e^{-\beta\hat{H}_{0}}\}=\prod_{i}\mathcal{Z}_{i}^{(0)},
~
\mathcal{Z}_{i}^{(0)} =\sum_{n=0}^{\infty}e^{-\beta f_{i}(n)}
\end{align}
is the partition function of the unperturbed system.

With the partition function $\mathcal{Z}$ in hand, the grand-canonical free energy $\mathcal{F}=-\beta^{-1}\ln\mathcal{Z}$ can be expressed as
\begin{equation}
\mathcal{F}=\mathcal{F}_{0}-\frac{1}{\beta}\ln\left[1+{\displaystyle \sum_{n=1}}\frac{\mathcal{Z}^{\left(n\right)}\left[j,j^{*}\right]}{\mathcal{Z}^{\left(0\right)}}\right],
\label{F-definiton}
\end{equation}
with $\mathcal{F}_{0}=-\beta^{-1}\ln\mathcal{Z}^{(0)}$ being the free energy of the unperturbed system.
Combining Eqs. (\ref{H1-j}), (\ref{Z_n}) and (\ref{F-definiton}) and by means of the formula $\ln (1+x)=x -\frac{1}{2}x^{2}+\cdots$, it is possible to directly expand $\mathcal{F}$ in a power series of perturbation parameters $j$, $j^{*}$ and $t$.
Note that such a complex yet straightforward calculation may allow one to make some interesting observations.
Firstly, the thermal averages in Eq.(\ref{Z_n}), thus the free energy, can be expressed in terms of $n$th thermal Green's functions with respect to the unperturbed system, which is defined by
\begin{equation}
  G_{n}^{(0)}\left(\tau_{1}^{\prime},i_{1}^{\prime};\ldots;\tau_{n}^{\prime},i_{n}^{\prime}|\tau_{1},i_{1};\ldots;\tau_{n},i_{n}\right)\equiv\left\langle \hat{T}\left[\hat{a}_{i_{1}^{\prime}}^{\dagger}\left(\tau_{1}^{\prime}\right)\hat{a}_{i_{1}}\left(\tau_{1}\right)\cdots\hat{a}_{i_{n}^{\prime}}^{\dagger}\left(\tau_{n}^{\prime}\right)\hat{a}_{i_{n}}\left(\tau_{n}\right)\right]\right\rangle _{0}.
\end{equation}
Meanwhile, it is also interesting to find that in the expression of $\mathcal{F}$ the coefficient of the 2nd-order term of $j$ ($j^{*}$), which plays an important role in the framework of Ginzburg-Landau theory, is exactly the 1-particle Green's function with respect to the original Hamiltonian $\hat{H}=\hat{H}_{0}+\hat{H}_{1}$,
\begin{equation}
  G_{1}\left(\tau^{\prime},i^{\prime}|\tau,i\right)=\frac{1}{\mathcal{Z}_{\hat{H}}}\mathrm{Tr}\left\{ e^{-\beta\hat{H}}\hat{T}\left[\hat{a}_{i^{\prime}}^{\dagger}\left(\tau^{\prime}\right)\hat{a}_{i}\left(\tau\right)\right]\right\}
  ,\label{G1}
\end{equation}
where $\mathcal{Z}_{\hat{H}}=\mathrm{Tr}\left\{ e^{-\beta\hat{H}}\right\} $ is the partition function of $\hat{H}$.
To be specific, we firstly pick up one term of $j$, one term of $j^{*}$ and $(n-2)$ terms of $t_{ij}$ in Eq. (\ref{Z_n}), then according to Eq. (\ref{F-definiton}) we obtain the second-order term of $j$ and $j^{*}$ in the expression of $\mathcal{F}$
\begin{align}
  [j^{2}]_{\mathcal{F}}=-\frac{1}{\beta}{\displaystyle \sum_{i^{\prime},i}}\int_{0}^{\beta}d\tau^{\prime}\int_{0}^{\beta}d\tau G_{i^{\prime},i}\left(\tau^{\prime},\tau\right)j_{i^{\prime}}\left(\tau^{\prime}\right)j_{i}^{*}\left(\tau\right),
\end{align}
where the coefficient takes the following form
\begin{align}
  G_{i^{\prime},i}\left(\tau^{\prime},\tau\right)=& {\displaystyle \sum_{n=0}}\frac{1}{n!}{\displaystyle \sum_{i_{1},j_{1},\cdots i_{n},j_{n}}}\int_{0}^{\beta}d\tau_{1}\dots\int_{0}^{\beta}d\tau_{n}t_{i_{1}j_{1}}\cdots t_{i_{n}j_{n}}
  \nonumber \\ &
  \times\left<\hat{T}\left[\hat{a}_{i^{\prime}}^{\dagger}\left(\tau^{\prime}\right)\hat{a}_{i}\left(\tau\right)\hat{a}_{i_{1}}^{\dagger}\left(\tau_{1}\right)\hat{a}_{j_{1}}\left(\tau_{1}\right)\dots\hat{a}_{i_{n}}^{\dagger}\left(\tau_{n}\right)\hat{a}_{j_{n}}\left(\tau_{n}\right)\right]\right>_{0}.
  \label{G_ij}
\end{align}
While expressing the 1-particle Green function $G_{1}\left(\tau^{\prime},i^{\prime}|\tau,i\right)$ defined by Eq. (\ref{G1}) in the imaginary time Dirac picture, one can easily check that it coincides with the second-order term coefficient $G_{i,j}\left(\tau_{1,}\tau_{2}\right)$ shown in Eq. (\ref{G_ij}).

\subsection{Cumulant expansion of free energy}

In principle, one could calculate the above thermal Green function straightforwardly and therefore give the expression of the free energy.
But the calculation will become extremely cumbersome when it goes to higher order.
Nevertheless, based on the locality of the unperturbed Hamiltonian,
an alternative approach has been developed to simplify such calculations, which is the so-called \emph{cumulant expansion} \cite{2009Bradlyn,2013MOhliger,1991Metzner-cumulant}.
According to the linked cluster theorem\cite{1991Metzner-cumulant,1990Gelfand-cumulant}, the free energy $\mathcal{F}$ can be expanded in terms of cumulants obtained from performing functional derivatives of a single generating functional with respect to the currents. Considering that the unperturbed  Hamiltonian $\hat{H}_{0}$ is a sum of local terms, the generating functional is defined as \cite{2009Bradlyn}
\begin{align}
C_{0}^{(0)}[j,j^{*}]=\ln\left<\hat{T}\exp\left[-\sum_{i}{\displaystyle \int_{0}^{\beta}\mathrm{d}\tau\left[j_{i}\left(\tau\right)\hat{a}_{i}^{\dagger}\left(\tau\right)+j_{i}^{*}\left(\tau\right)\hat{a}_{i}\left(\tau\right)\right]}\right]\right>_{0},
 \label{C0}
\end{align}
and the cumulants are defined by
\begin{align}
C_{2n}^{(0)}(i_{1}^{\prime},\tau_{1}^{\prime};\dots;i_{n}^{\prime},\tau_{n}^{\prime}|i_{1},\tau_{1};\dots;i_{n},\tau_{n})=\left.\frac{\delta^{2n}C_{0}^{(0)}[j,j^{*}]}{\delta j_{i'_{1}}(\tau'_{1})\dots\delta j_{i'_{n}}(\tau'_{n})\delta j_{i_{1}}^{*}(\tau_{1})\dots\delta j_{i_{n}}^{*}(\tau_{n})}\right|_{j=j^{*}=0}.
\label{C2n}
\end{align}
Note that because of the local structure of the generating functional Eq. (\ref{C0}), the cumulants $C_{2n}^{(0)}$ also become local, i.e.
\begin{equation}
C_{2n}^{(0)}(i_1',\tau_1';\dots;i_n',\tau_n'| i_1,\tau_1;\dots;i_n,\tau_n)
={_{i_1}C}_{2n}^{(0)}(\tau_1',\dots,\tau_n'| \tau_1,\dots,\tau_n)\prod_{n,m}{\delta_{i'_n,i_m}}.
\label{c2n-i}
\end{equation}
To be specific, we give the expressions of the first two cumulants according to the above definitions
\begin{align}
_iC^{(0)}_2(\tau_1|\tau_2)=&\left<\hat{T}\left[\hat{a}^{\dag}_i(\tau_1)\hat{a}_i(\tau_2)\right]\right>_0,
\label{c2_0}
\\
_iC^{(0)}_4(\tau_1,\tau_2|\tau_3,\tau_4)=&\left<\hat{T}\left[\hat{a}^{\dag}_i(\tau_1)\hat{a}^{\dag}_i(\tau_2)\hat{a}_i(\tau_3)
\hat{a}_i(\tau_4)\right]\right>_0 \nonumber \\
&-{_iC}^{(0)}_2(\tau_1|\tau_3)_iC^{(0)}_2(\tau_2|\tau_4)-_iC^{(0)}_2(\tau_1|\tau_4)_iC^{(0)}_2(\tau_2|\tau_3). \label{c4_0}
\end{align}

By means of cumulant decompositions \cite{2009Bradlyn,2012CNietner-JCM}, we have the chance to give the following perturbative expression of the free energy $\mathcal{F}$
\begin{align}
\mathcal{F}=& F_{0}-\frac{1}{\beta}\sum_{ij}\int_{0}^{\beta}d\tau_{1}\int_{0}^{\beta}d\tau_{2}\left[a_{2}^{(0)}(i,\tau_{1}|i,\tau_{2})\delta_{i,j}+a_{2}^{(1)}(i,\tau_{1}|j,\tau_{2})t_{ij}\right.
 \nonumber \\ &
 +\left.\sum_{k}a_{2}^{(2)}(i,\tau_{1}|j,\tau_{2})t_{ik}t_{kj}+\sum_{k}a_{2}^{(2)}(i,\tau_{1}|i,\tau_{2})t_{ik}t_{kj}\delta_{i,j}+\cdots\right]j_{i}(\tau_{1})j_{j}^{*}(\tau_{2})
 \nonumber \\ &
 -\frac{1}{4\beta}\sum_{ijkl}\int_{0}^{\beta}d\tau_{1}\int_{0}^{\beta}d\tau_{2}\int_{0}^{\beta}d\tau_{3}\int_{0}^{\beta}d\tau_{4}\left[a_{4}^{(0)}(i,\tau_{1};i,\tau_{2}|i,\tau_{3};i,\tau_{4})\delta_{i,j}\delta_{j,k}\delta_{k,l}\right.
 \nonumber \\ &
 +\left.2t_{ik}a_{4}^{(1)}(i,\tau_{1};i,\tau_{2}|k,\tau_{3};i,\tau_{4})\delta_{i,j}\delta_{i,l}+2t_{ij}a_{4}^{(1)}(i,\tau_{1};j,\tau_{2}|i,\tau_{3};i,\tau_{4})\delta_{i,k}\delta_{i,l}+\cdots\right]
 \nonumber \\ &
 \times j_{i}(\tau_{1})j_{j}(\tau_{2})j_{k}^{*}(\tau_{3})j_{l}^{*}(\tau_{4})+\mathcal{O}\left(j^{6}\right)
 ,\label{F}
\end{align}
where the expansion coefficients $a_{2n}^{(m)}$ are closely related to the cumulants and are defined by
\begin{align}
a_{2}^{(0)}(i,\tau_{1}|i,\tau_{2})=& _{i}C_{2}^{(0)}(\tau_{1}|\tau_{2}),
\label{a2-0}
\\
a_{2}^{(1)}(i,\tau_{1}|j,\tau_{2})=& \int_{0}^{\beta}d\tau^{\prime}_{i}C_{2}^{(0)}(\tau_{1}|\tau^{\prime})_{j}C_{2}^{(0)}(\tau^{\prime}|\tau_{2}),
\\
a_{2}^{(2)}(i,\tau_{1}|j,\tau_{2})=& \int_{0}^{\beta}d\tau_{1}^{\prime}\int_{0}^{\beta}d\tau_{2}^{\prime}{}_{i}C_{2}^{(0)}(\tau_{1}|\tau_{1}^{\prime})_{k}C_{2}^{(0)}(\tau_{1}^{\prime}|\tau_{2}^{\prime})_{j}C_{2}^{(0)}(\tau_{2}^{\prime}|\tau_{2}),
\\
a_{2}^{(2)}(i,\tau_{1}|i,\tau_{2})=& \int_{0}^{\beta}d\tau_{1}^{\prime}\int_{0}^{\beta}d\tau_{2}^{\prime}{}_{i}C_{4}^{(0)}(\tau_{1},\tau_{1}^{\prime}|\tau_{2}^{\prime},\tau_{2})_{k}C_{2}^{(0)}(\tau_{2}^{\prime}|\tau_{1}^{\prime})
,\label{a2-2}
\\
a_{4}^{(0)}(i,\tau_{1};i,\tau_{2}|i,\tau_{3};i,\tau_{4})=& {}_{i}C_{4}^{(0)}(\tau_{1},\tau_{2}|\tau_{3},\tau_{4}),
\label{a4_0-def}
\\
a_{4}^{(1)}(i,\tau_{1};i,\tau_{2}|k,\tau_{3};i,\tau_{4})=& \int_{0}^{\beta}d\tau^{\prime}{}_{i}C_{4}^{(0)}(\tau_{1},\tau_{2}|\tau^{\prime},\tau_{4})_{k}C_{2}^{(0)}(\tau^{\prime}|\tau_{3}),
\\
a_{4}^{(1)}(i,\tau_{1};j,\tau_{2}|i,\tau_{3};i,\tau_{4})=& \int_{0}^{\beta}d\tau^{\prime}_{j}C_{2}^{(0)}(\tau_{2}|\tau^{\prime})_{i}C_{4}^{(0)}(\tau_{1},\tau^{\prime}|\tau_{3},\tau_{4}).
\end{align}

Now, the calculation of coefficients $a_{2n}^{(m)}$ becomes the key to determine the free energy. In fact these coefficients can often be further simplified by going to Matsubara frequency space, where the integrations over time variables will amount to simple multiplications. Let us use the following conventions,
\begin{align}
\begin{cases}
g\left(\omega_{m}\right)=\frac{1}{\sqrt{\beta}}\int_{0}^{\beta}d\tau e^{i\omega_{m}\tau}g\left(\tau\right)\\
g\left(\tau\right)=\frac{1}{\sqrt{\beta}}{\displaystyle \sum_{m=-\infty}^{+\infty}g\left(\omega_{m}\right)e^{-i\omega_{m}\tau}}
\end{cases}
\label{matsubara-convention}
\end{align}
where $\omega_m =\frac{2\pi m}{\beta},~ m\in\mathbb{Z}$ is the Matsubara frequency with $\text{\ensuremath{\beta}=\ensuremath{\frac{1}{k_{B}T}}}$.
Due to the time-translation invariancy of the unperturbed Hamiltonian, it is easy to verify that the cumulants (and thus $a_{2n}^{(m)}$) only depend on time interval. Such a time invariance indicates that the Matsubara frequency conservation must hold.
Combining Eqs. (\ref{c2_0}), (\ref{a2-0}) with the Matsubara transformation (\ref{matsubara-convention}) we find
\begin{align}
a_2^{(0)}(i,\omega_{m1}|i,\omega_{m2})&=a_2^{(0)}(i,\omega_{m1})\delta_{\omega_{m1},\omega_{m2}}, \label{a2_0-w}
\end{align}
Similarly, we have
\begin{align}
a_2^{(1)}(i,\omega_{m1}|j,\omega_{m2})&=a_2^{(0)}(i,\omega_{m1})a_2^{(0)}(j,\omega_{m1})\delta_{\omega_{m2},\omega_{m2}},
\\
a_{2}^{(2)}(i,\omega_{m1}|j,\omega_{m2})&=a_{2}^{(0)}(i,\omega_{m1})a_{2}^{(0)}(k,\omega_{m2})a_{2}^{(0)}(j,\omega_{m2})\delta_{\omega_{m1},\omega_{m2}},
\\
a_4^{(0)} (i,\omega_{m1};i,\omega_{m2}|i,\omega_{m3};i,\omega_{m4})&=a_4^{(0)} (i,\omega_{m1};i,\omega_{m2}|i,\omega_{m4})
\delta_{\omega_{m1}+\omega_{m2},\omega_{m3}+\omega_{m4}},
\label{a4_0-w}
\\
a_4^{(1)}(i,\omega_{m1};i,\omega_{m2}|j,\omega_{m3};i,\omega_{m4})&=a_4^{(0)}(i,\omega_{m1};i,\omega_{m2}|i,\omega_{m4})
a_2^{(0)}(j,\omega_{m3})\delta_{\omega_{m1}+\omega_{m2},\omega_{m3}+\omega_{m4}},
\\
a_{4}^{(1)}(i,\omega_{m1};j,\omega_{m2}|i,\omega_{m3};i,\omega_{m4})&=a_{4}^{(0)}(i,\omega_{m1}|i,\omega_{m3};i,\omega_{m4})a_{2}^{(0)}(j,\omega_{m2})\delta_{\omega_{m1}+\omega_{m2},\omega_{m3}+\omega_{m4}}.
\end{align}
Hence $a_{2n}^{(0)}$ become the basic blocks.
According to Eqs. (\ref{c2_0}), (\ref{a2-0}) and (\ref{a2_0-w}), a straightforward calculation gives the expression of $a_{2}^{(0)}$ as
\begin{align}
a_2^{(0)}(i,\omega_m)=\frac{1}{\mathcal{Z}^{(0)}}\sum_{n=0}^{\infty}e^{-\beta f_i(n)} \left[\frac{n+1}{f_i(n+1)-f_i(n)-i\omega_{m}}-\frac{n}{f_i(n)-f_i(n-1)-i\omega_{m}}\right]
.\label{a2}
\end{align}
As for the fourth-order coefficient $a_{4}^{(0)}$, we give a detailed calculation and its lengthy expression in Appendix A.
Unlike other coefficients, the second-order correction coefficient $a_{2}^{(2)}(i,\omega_{m1}|i,\omega_{m2})$ cannot decompose into simple product of $a_{2n}^{(0)}$, and the result is presented in Appendix B due to its complexity and length.
Finally, inserting the Matsubara expansions of $\hat{a}(\tau)$, $\hat{a}^{\dagger}(\tau)$, $j(\tau)$ and $j^{*}(\tau)$ into Eq. (\ref{F}) gives the free energy in Matsubara representation
\begin{align}
\mathcal{F}= & F_{0}-\frac{1}{\beta}\Big[\sum_{ij}\sum_{\omega_{m1},\omega_{m2}}G_{ij}(\omega_{m1},\omega_{m2})j_{i}(\omega_{m1})j_{j}^{*}(\omega_{m2})
\nonumber \\ &
+\left.\sum_{ijkl}\sum_{\substack{\omega_{m1},\omega_{m2}\\
\omega_{m3},\omega_{m4}
}
}H_{ijkl}(\omega_{m1},\omega_{m2},\omega_{m3},\omega_{m4})j_{i}(\omega_{m1})j_{j}(\omega_{m2})j_{k}^{*}(\omega_{m3})j_{l}^{*}(\omega_{m4})\right]+\mathcal{O}\left(j^{6}\right), \label{F-GH}
\end{align}
where we have used the following abbreviations
\begin{align}
G_{ij}(\omega_{m1},\omega_{m2})\equiv & G_{ij}(\omega_{m1})\delta_{\omega_{m1},\omega_{m2}},
\nonumber \\
G_{ij}(\omega_{m1})=& \left[a_{2}^{(0)}(i,\omega_{m1})\delta_{i,j}+a_{2}^{(0)}(i,\omega_{m1})a_{2}^{(0)}(j,\omega_{m1})t_{ij}\right.
\nonumber \\ &
+\sum_{k}a_{2}^{(0)}(i,\omega_{m1})a_{2}^{(0)}(k,\omega_{m2})a_{2}^{(0)}(j,\omega_{m2})t_{ik}t_{kj}
\nonumber \\ &
\left.+\sum_{k}a_{2}^{(2)}(i,\omega_{m1}|i,\omega_{m2})t_{ik}t_{kj}\delta_{i,j}+\mathcal{O}\left(t^{3}\right)\right]
,\label{G-ij}
\\
H_{ijkl}(\omega_{m1},\omega_{m2},\omega_{m3},\omega_{m4})=& \frac{\delta_{\omega_{m1}+\omega_{m2},\omega_{m3}+
\omega_{m4}}}{4}a_4^{(0)}(i,\omega_{m1};i,\omega_{m2}|i,\omega_{m4})\Big\{\delta_{ij}\delta_{jk}\delta_{kl}\Big. \nonumber \\ &
+\left.2\left[a_{2}^{(0)}(k,\omega_{m3})t_{ik}\delta_{ij}\delta_{il}+a_{2}^{(0)}(j,\omega_{m2})t_{ij}\delta_{ik}\delta_{il}\right]+\mathcal{O}\left(t^{2}\right)\right\} .
\label{H-ijkl}
\end{align}

\subsection{Effective action of disordered BHM}
While dealing with quenched-disorder problem, the disorder ensemble average should be taken after the thermodynamic ensemble average \cite{1986BinderYoung}. Therefore the disorder-averaged free energy in Matsubara space needs to be given first
\begin{align}
  \mathcal{\overline{F}}=&\overline{F_{0}}-\frac{1}{\beta}\left[\sum_{ij}\sum_{\omega_{m1}}\overline{G_{ij}(\omega_{m1})}j_{i}(\omega_{m1})j_{j}^{*}(\omega_{m1})\right.
 \nonumber \\ &
  +\left.\sum_{ijkl}\sum_{\substack{\omega_{m1},\omega_{m2}\\
\omega_{m3},\omega_{m4}
}
}\overline{H_{ijkl}(\omega_{m1},\omega_{m2},\omega_{m3},\omega_{m4})}j_{i}(\omega_{m1})j_{j}(\omega_{m2})j_{k}^{*}(\omega_{m3})j_{l}^{*}(\omega_{m4})+\overline{\mathcal{O}\left(j^{6}\right)}\right],
 \label{F-disorder}
\end{align}
where
\begin{align}
  \overline{G_{ij}(\omega_{m1})}=& \left[\overline{a_{2}^{(0)}(i,\omega_{m1})}\delta_{i,j}+\overline{a_{2}^{(0)}(i,\omega_{m1})}^{2}t_{ij}+\sum_{k}\overline{a_{2}^{(0)}(i,\omega_{m1})}^{3}t_{ik}t_{kj}\right.
  \label{G-ij-disorder} \\ &
  \left.+\sum_{k}\overline{a_{2}^{(2)}(i,\omega_{m1}|i,\omega_{m2})}t_{ik}t_{kj}\delta_{i,j}+\mathcal{O}\left(t^{3}\right)\right],
  \nonumber \\
  \overline{H_{ijkl}(\omega_{m1},\omega_{m2},\omega_{m3},\omega_{m4})}=& \frac{\delta_{\omega_{m1}+\omega_{m2},\omega_{m3}+\omega_{m4}}}{4}\overline{a_{4}^{(0)}(i,\omega_{m1};i,\omega_{m2}|i,\omega_{m4})}\left\{ \delta_{ij}\delta_{jk}\delta_{kl}\right.
  \nonumber \\ &
  +\left.2\left[\overline{a_{2}^{(0)}(k,\omega_{m3})}t_{ik}\delta_{ij}\delta_{il}+\overline{a_{2}^{(0)}(j,\omega_{m2})}t_{ij}\delta_{ik}\delta_{il}\right]+\mathcal{O}\left(t^{2}\right)\right\} .
  \label{H-ijkl-disorder}
\end{align}
Note that the disorder is assumed to be spatially uncorrelated, thus the coefficients $a_{2}^{(0)}$ of different sites become also uncorrelated, i.e.
\begin{equation}
  \overline{a_{2}^{(0)}(i,\omega_{m1})a_{2}^{(0)}(j,\omega_{m1})}=\overline{a_{2}^{(0)}(i,\omega_{m1})}^{2}
.\end{equation}

As explored in the context of thermal phase transitions\cite{2001Kleinert,2002Justin} and further developed while dealing with quantum phase transitions in clean lattice systems\cite{2009Santos,2009Bradlyn},
we introduce the following disorder-averaged parameter fields
\begin{align}
\overline{\psi_{i}(\omega_{m})}=\overline{\left<\hat{a}_{i}(\omega_{m})\right>}=\beta\frac{\delta\overline{\mathcal{F}}}{\delta j_{i}^{*}(\omega_{m})},
\nonumber \\
\overline{\psi_{i}(\omega_{m})}^{*}=\overline{\left<\hat{a}_{i}^{\dagger}(\omega_{m})\right>}=\beta\frac{\delta\overline{\mathcal{F}}}{\delta j_{i}(\omega_{m})}.
\label{psi-definition}
\end{align}
The above equations motivate one to perform a Legendre transformation of $\overline{\mathcal{F}}$, which gives us the following Ginzburg-Landau effective action
\begin{equation}
\Gamma\left[\overline{\psi_{i}},\overline{\psi_{i}}^{*}\right]=\overline{\mathcal{F}}-\frac{1}{\beta}\sum_{i}\sum_{\omega_{m}}\left[\overline{\psi_{i}(\omega_{m})}j_{i}^{*}(\omega_{m})+\overline{\psi_{i}(\omega_{m})}^{*}j_{i}(\omega_{m})\right].
 \label{Gamma}
\end{equation}
From the above expression it is obvious that
\begin{align}
j_{i}(\omega_{m})=-\beta\frac{\delta\Gamma}{\delta\overline{\psi_{i}(\omega_{m})}^{*}},
\nonumber \\
j_{i}^{*}(\omega_{m})=-\beta\frac{\delta\Gamma}{\delta\overline{\psi_{i}(\omega_{m})}}.
\label{j-definition}
\end{align}

Remember that in real physical situation, the artificially introduced currents should vanish, i.e. in the end we should set $j_{i}(\omega_{m})=j_{i}^{*}(\omega_{m})=0$, meaning that
\begin{equation}
\frac{\delta\Gamma}{\delta\overline{\psi_{i}(\omega_{m})}^{*}}=\frac{\delta\Gamma}{\delta\overline{\psi_{i}(\omega_{m})}}=0.
\label{gamma-equilibrium}
\end{equation}
The above equilibrium condition means that the effective action $\Gamma$ is stationary with respect to the parameter field $\psi$, $\psi^{*}$.
To obtain the Ginzburg-Landau expansion of the effective action, we need to express the current $j$, $j^{*}$ in terms of $\psi$, $\psi^{*}$ first.
According to Eqs. (\ref{F-disorder}), (\ref{psi-definition}), we have
\begin{align}
  \overline{\psi_{p}(\omega_{m})}=& -\sum_{i}\overline{G_{ip}(\omega_{m})}j_{i}(\omega_{m})
  \nonumber \\ &
  -2\sum_{ijk}\sum_{\substack{\omega_{m1},\omega_{m2}\\
\omega_{m3}
}
}\overline{H_{ijkp}(\omega_{m1},\omega_{m2},\omega_{m3},\omega_{m})}j_{i}(\omega_{m1})j_{j}(\omega_{m2})j_{k}^{*}(\omega_{m3})+\mathcal{O}\left(j^{5}\right)
,\nonumber \\
\overline{\psi_{p}^{*}(\omega_{m})}=& -\sum_{j}\overline{G_{pj}(\omega_{m})}j_{j}^{*}(\omega_{m})
\nonumber \\ &
-2\sum_{ikl}\sum_{\substack{\omega_{m1}\\
\omega_{m3},\omega_{m4}
}
}\overline{H_{ipkl}(\omega_{m1},\omega_{m},\omega_{m3},\omega_{m4})}j_{i}(\omega_{m1})j_{k}^{*}(\omega_{m3})j_{l}^{*}(\omega_{m4})+\mathcal{O}\left(j^{5}\right)
.\label{psi-mn}
\end{align}
Thus the lowest order approximation of $j_{i}(\omega_{m})$ can be written as
\begin{align}
j_{i}(\omega_{m})\simeq J_{i}(\omega_{m})=-\sum_{p}\left[\overline{G_{pi}(\omega_{m})}\right]^{-1}\overline{\psi_{p}(\omega_{m})},
 \nonumber \\
j_{j}^{*}(\omega_{m})\simeq J_{j}^{*}(\omega_{m})=-\sum_{q}\left[\overline{G_{jq}(\omega_{m})}\right]^{-1}\overline{\psi_{q}^{*}(\omega_{m})},
 \label{j-J}
\end{align}
where $\overline{G^{-1}(\omega_{m})}$ is the inverse of $\overline{G(\omega_{m})}$, i.e. ${\displaystyle \sum_{p}}\overline{G_{ip}(\omega_{m})}\left[\overline{G_{pj}(\omega_{m})}\right]^{-1}=\delta_{i,j}
 $.
Now inverting (\ref{psi-mn}) recursively with (\ref{j-J}) gives
\begin{align}
  j_{i}(\omega_{m})=&-\sum_{p}\left[\overline{G_{pi}(\omega_{m})}\right]^{-1}\left[\overline{\psi_{p}(\omega_{m})}\right.
 \nonumber \\ &
  \left.-2\sum_{pjk}\sum_{\omega_{m1},\omega_{m2},\omega_{m3}}\overline{H_{pjki}(\omega_{m1},\omega_{m2},\omega_{m3},\omega_{m})}J_{p}(\omega_{m1})J_{j}(\omega_{m2})J_{k}^{*}(\omega_{m3})\right]+\mathcal{O}\left(J^{5}\right),
  \nonumber \\
  j_{j}^{*}(\omega_{m})=& -\sum_{p}\left[\overline{G_{jp}(\omega_{m})}\right]^{-1}\left[\overline{\psi_{p}(\omega_{m})}^{*}\right.
  \nonumber \\ &
  \left.+\sum_{ikl}\sum_{\substack{\omega_{m1}\\
\omega_{m3},\omega_{m4}
}
}\overline{H_{ipkl}(\omega_{m1},\omega_{m},\omega_{m3},\omega_{m4})}J_{i}(\omega_{m1})J_{k}^{*}(\omega_{m3})J_{l}^{*}(\omega_{m4})\right]+\mathcal{O}\left(J^{5}\right).
\end{align}
Combining the above expression of $j$ with Eqs. (\ref{F-disorder}), (\ref{Gamma}) gives a general expression of effective action
\begin{align}
  \Gamma[\psi_{i},\psi_{i}^{*}]=&\overline{F_{0}}+\frac{1}{\beta}\sum_{ij}\sum_{\omega_{m1},\omega_{m2}}\left[\overline{G_{ij}(\omega_{m1},\omega_{m2})}\right]^{-1}\overline{\psi_{i}(\omega_{m1})}\cdot\overline{\psi_{j}(\omega_{m2})}^{*}
  \nonumber \\ &
-\frac{1}{\beta}\sum_{ijkl}\sum_{\substack{\omega_{m1},\omega_{m2}\\
\omega_{m3},\omega_{m4}
}
}\overline{H_{ijkl}^{\prime}(\omega_{m1},\omega_{m2},\omega_{m3},\omega_{m4})}\cdot\overline{\psi_{i}(\omega_{m1})}\cdot\overline{\psi_{j}(\omega_{m2})}\cdot\overline{\psi_{k}(\omega_{m3})}^{*}\cdot\overline{\psi_{l}(\omega_{m4})}^{*}
 \label{gamma-mn}
\end{align}
Here,  $\left[\overline{G(\omega_{m1},\omega_{m2})}\right]^{-1}=\left[\overline{G(\omega_{m1})}\right]^{-1}\delta_{\omega_{m1},\omega_{m2}}$ is the inverse of $\overline{G(\omega_{m1},\omega_{m2})}$, which is the disorder-averaged single-particle Green function with respect to Hamiltonian $\hat{H}$ in Matsubara space according to Eqs. (\ref{G_ij}), (\ref{matsubara-convention}).

\section{Phase Boundary}

In the following we focus on the effective action of disordered Bose-Hubbard model at zero-temperature limit.
In static case, the order parameter fields will become constant in time (thus Matsubara frequency). Meanwhile, after the disorder ensemble average they should also become independent of space, i.e. $\overline{\psi_{i}(\omega_{m})}=\overline{\psi_{j}(\omega_{m})}=\overline{\psi(\omega_{m})}$. Therefore at the equilibrium situation, we have
\begin{align}
  \overline{\psi_{i}(\omega_{m})}=\sqrt{\beta}\cdot\overline{\psi}\delta_{\omega_{m},0}.
\end{align}
Using this ansatz we find that the effective action in Eq. (\ref{gamma-mn}) takes the following form
\begin{align}
  \Gamma\left[\overline{\psi},\overline{\psi}^{*}\right]=\overline{F_{0}}+\sum_{ij}\left(\overline{G_{ij}}\right)^{-1}\left|\overline{\psi}\right|^{2}-\sum_{ijkl}\beta\overline{H_{ijkl}^{\prime}}\left|\overline{\psi}\right|^{4}+\mathcal{O}(\left|\psi\right|^{6}),
 \label{gamma-0T}
\end{align}
where
\begin{align}
  \overline{G_{ij}}=\overline{a_{2}^{(0)}(i,0)}\delta_{i,j}+\overline{a_{2}^{(0)}(i,0)}^{2}t_{ij}+\sum_{k}\overline{a_{2}^{(0)}(i,0)}^{3}t_{ik}t_{kj}+\sum_{k}\overline{a_{2}^{(2)}(i,0|i,0)}t_{ik}t_{kj}\delta_{i,j}+\overline{\mathcal{O}\left(t^{3}\right)}.
\end{align}
Note that the 2nd-order term coefficient $\overline{G_{ij}}$ has been approximated to second order in the tunneling, which can be seen later that it plays an essential role in determining the beyond mean-field phase boundary.

By now, one can see that the effective action given above takes the form of $\phi^{4}$ theory \cite{1970Landau}. As long as the coefficient of $\left|\overline{\psi}\right|^{4}$ is positive, the boundary of phase with $\left|\psi\right|=0$ and phase with $\left|\psi\right|\neq0$ can be given.
We remark here that unlike the clean case, where both the order parameter $\psi$ and its modulus $\left|\psi\right|$ can serve as the SF order parameter, in disordered system $\overline{\psi}$ and $\left|\overline{\psi}\right|$ will have different behaviors. When disorder is introduced, the complex number $\overline{\psi}$ is nonzero only in SF phase and vanishes in BG due to the disorder average over its phase. Whereas $\left|\overline{\psi}\right|$ goes to zero only in MI and survives in BG as the macroscopically nonzero SF puddles appears when the system comes into the BG phase.
Therefore, we argue here that the modulus of the parameter $\left|\overline{\psi}\right|$ in our case can be used to distinguish the MI and BG within the framework of $\phi^{4}$ theory, where $\overline{\psi}$ and $\overline{\psi}^{*}$ appear in pairs and thus only the information of $\left|\overline{\psi}\right|$ can be given.

\subsection{Mean-field result}
\begin{figure}[H]
\centering\includegraphics[width=10cm]{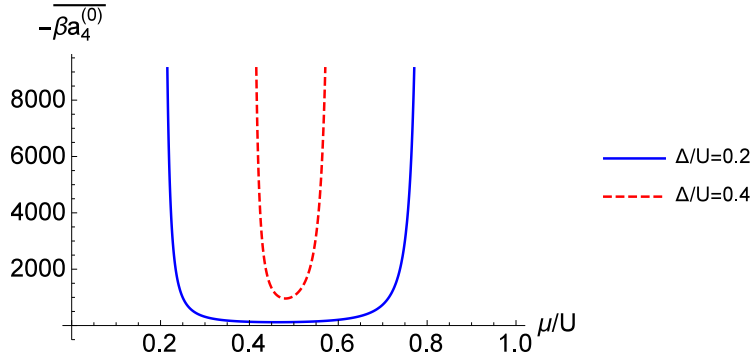}
\caption{The graph of the coefficient of $\left|\overline{\psi}\right|^{4}$ in the case of $n=1$, $\Delta/U=0.2$ (blue line) and $\Delta/U=0.4$ (red line).}
\label{coefficients-c4MF}
\end{figure}

\begin{figure}[H]
\centering\includegraphics[width=7cm]{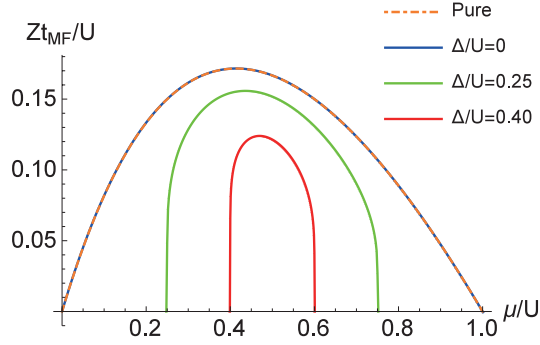}~~
\caption{MI-BG phase diagram at different disorder strength $\Delta/U$ in the mean-field limit with the filling factor $n=1$. The orange dotdashed line is the result obtained in clean lattice system \cite{2009Santos}.}
\label{phase-diagram-mf}
\end{figure}

Keeping the terms only up to the first order of tunneling $t_{ij}$, the mean-field effective action in zero temperature limit reads
\begin{align}
  \Gamma_{MF}\left[\overline{\psi},\overline{\psi}^{*}\right]=\overline{F_{0}}+N_{S}\left[\left(\frac{1}{\overline{a_{2}^{(0)}(i,0)}}-\sum_{j}t_{ij}\right)\left|\overline{\psi}\right|^{2}-\frac{\beta\overline{a_{4}^{(0)}(i,0)}}{4\left(\overline{a_{2}^{(0)}(i,0)}\right)^{4}}\left|\overline{\psi}\right|^{4}\right]+\mathcal{O}(\overline{\left|\psi\right|}^{6}),
 \label{gamma-mean-field}
\end{align}
where $N_{S}$ is the total site number. The coefficients $a_{2}^{(0)}$ and $\beta a_{4}^{(0)}$ can be obtained from Eqs. (\ref{fn}), (\ref{a2}) and (\ref{a4})
\begin{align}
  a_{2}^{\left(0\right)}\left(i,0\right)=&\frac{n+1}{nU-\mu_{i}}-\frac{n}{\left(n-1\right)U-\mu_{i}},
  \\
  \beta a_{4}^{(0)}(i,0)=& 4n\left(n-1\right)\left[\frac{1}{U}{\displaystyle \frac{1}{\left(\left(n-1\right)U-\mu_{i}\right)^{2}}}+\frac{2}{U^{2}}{\displaystyle \frac{1}{\left(\left(n-1\right)U-\mu_{i}\right)}-\frac{2}{U^{2}}}{\displaystyle \frac{1}{\left(n-\frac{3}{2}\right)U-\mu_{i}}}\right]
  \nonumber \\ &
  -4n\left(n+1\right)\left[\frac{1}{U}\frac{1}{\left(\left(n-1\right)U-\mu_{i}\right)^{2}}+\frac{1}{U}\frac{1}{\left(nU-\mu_{i}\right)^{2}}-\frac{2}{U^{2}}\frac{1}{\left(\left(n-1\right)U-\mu_{i}\right)}+\frac{2}{U^{2}}\frac{1}{\left(nU-\mu_{i}\right)}\right]
  \nonumber \\ &
  +4\left(n+1\right)\left(n+2\right)\left[\frac{1}{U}\frac{1}{\left(nU-\mu_{i}\right)^{2}}-\frac{2}{U^{2}}\frac{1}{\left(nU-\mu_{i}\right)}+\frac{2}{U^{2}}\frac{1}{\left(\left(n+\frac{1}{2}\right)U-\mu_{i}\right)}\right]
  \nonumber \\ &
  -\frac{4\left(n+1\right)^{2}}{\left(nU-\mu_{i}\right)^{3}}+\frac{4n^{2}}{\left(\left(n-1\right)U-\mu_{i}\right)^{3}}.
\end{align}
After the disorder ensemble average, we have \begin{align}
  \overline{a_{2}^{\left(0\right)}\left(i,0\right)}=\frac{1}{2\Delta}\left[n\ln\frac{\left(n-1\right)U-\mu-\Delta}{\left(n-1\right)U-\mu+\Delta}-\left(n+1\right)\ln\frac{nU-\mu-\Delta}{nU-\mu+\Delta}\right].
\label{a2-avg}
\end{align}
Since we only care about the sign of the coefficient of $\left|\overline{\psi}\right|^{4}$, for brevity we just show the graph of the disorder-averaged coefficient $-\beta\overline{a_{4}^{(0)}(i,0)}$ in Fig. \ref{coefficients-c4MF},
From the above figure we can see that the coefficient of $\left|\overline{\psi}\right|^{4} $ in Eq. (\ref{gamma-mean-field}) must be positive, indicating the second-order phase transition happens here\cite{1970Landau}.
Thus the phase boundary can be obtained by vanishing the coefficient of $\left|\overline{\psi}\right|^{2} $ in Eq. (\ref{gamma-mean-field}),
\begin{align}
  Zt_{MF}=\frac{1}{\overline{a_{2}^{(0)}(i,0)}},
  \label{t-MF}
\end{align}
where $Z=2d$ is the number of nearest neighbor of a given site.
The above phase boundary equation in mean-field limit gives us the phase diagram showed in Fig. \ref{phase-diagram-mf}.
Clearly, when the disorder vanishes our result coincides with that obtained in clean lattice system \cite{2009Santos}. As disorder grows, the MI region will be suppressed as expected. Our phase boundary results shown here coincide with the previous mean-field theory \cite{1989Fisher,2006Krutitsky}.

\subsection{Beyond mean field}
One of the advantages of GLT showed here is that the effects of quantum fluctuations beyond mean-field can be incorporated. To the second order of the hopping parameter $t$, the coefficient of $\left|\overline{\psi}\right|^{2}$ is
\begin{align}
  \left(\overline{G_{ij}}\right)^{-1}=\frac{1}{\overline{a_{2}^{(0)}(i,0)}}\delta_{i,j}-t_{ij}-\sum_{k}\frac{\overline{a_{2}^{(2)}(i,0|i,0)}}{\overline{a_{2}^{(0)}(i,0)}^{2}}t_{ik}t_{ki}\delta_{i,j}.
\end{align}
In such a case, the effective action takes the following form
\begin{align}
  \Gamma\left[\overline{\psi},\overline{\psi}^{*}\right]=\overline{F_{0}}+N_{S}\left[\frac{1}{\overline{a_{2}^{(0)}(i,0)}}-Zt-\frac{\overline{a_{2}^{(2)}(i,0|i,0)}}{\overline{a_{2}^{(0)}(i,0)}^{2}}Zt^{2}\right]\left|\overline{\psi}\right|^{2}-\sum_{ijkl}\beta\overline{H_{ijkl}^{\prime}}\left|\overline{\psi}\right|^{4}+\mathcal{O}(\left|\psi\right|^{6})
.\end{align}

We assume that higher order corrections would not change the class of phase transition, i.e. the vanishing of the coefficient of $\left|\overline{\psi}\right|^{2}$ still gives the phase boundary. Hence the phase boundary equation can be obtained from solving the following equation
\begin{align}
  \frac{1}{\overline{a_{2}^{(0)}(i,0)}}-Zt_{BMF}-\frac{\overline{a_{2}^{(2)}(i,0|i,0)}}{\overline{a_{2}^{(0)}(i,0)}^{2}}Zt_{BMF}^{2}=0,
\end{align}
which gives the following meaningful solution
\begin{align}
  Zt_{BMF}=\frac{1}{2}\frac{Z\cdot\overline{a_{2}^{(0)}(i,0)}^{2}}{\overline{a_{2}^{(2)}(i,0|i,0)}}\left[\sqrt{1+4\frac{\overline{a_{2}^{(2)}(i,0|i,0)}}{Z\cdot\overline{a_{2}^{(0)}(i,0)}^{3}}}-1\right].
\end{align}

\begin{figure}[H]
\centering\includegraphics[width=7cm]{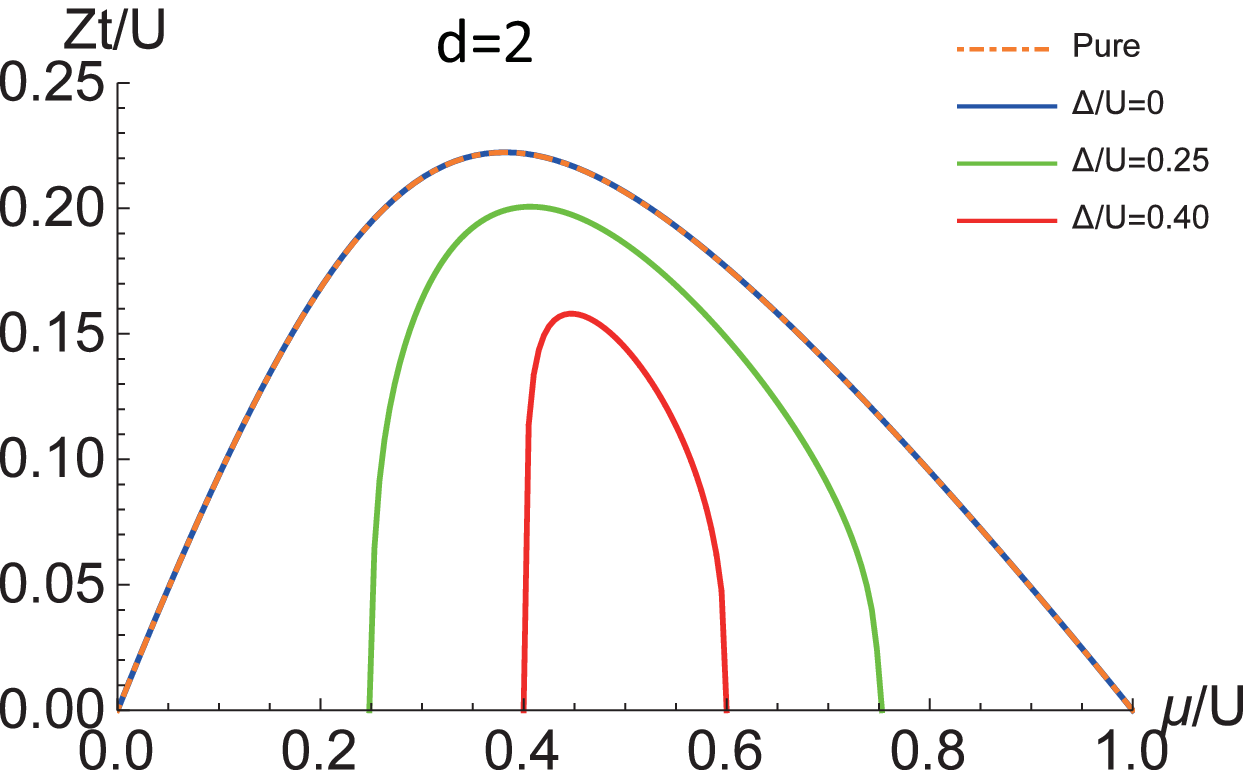}~~
\centering\includegraphics[width=7cm]{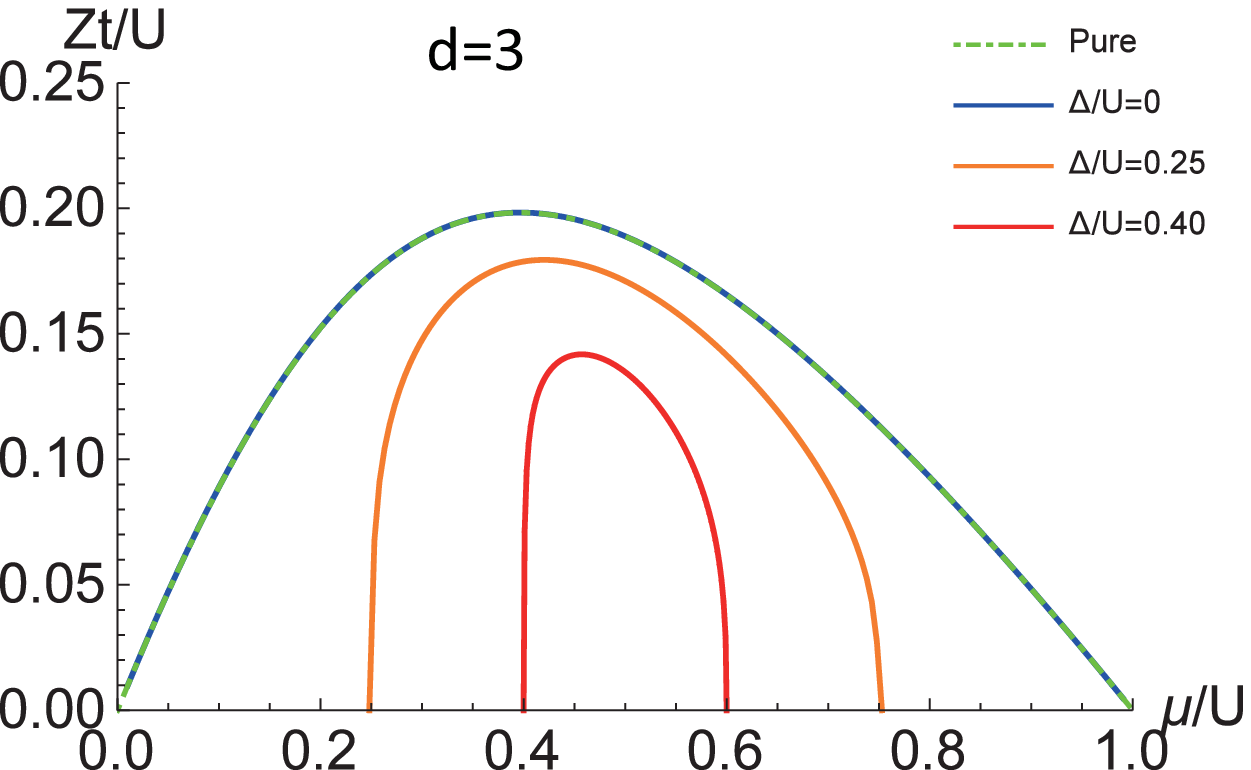}
\caption{Beyond mean-field phase diagrams of MI-BG of disordered BHM at zero temperature limit in $d=2$ (left) and $d=3$ (right) dimensions. The dot-dashed lines are the results obtained in clean lattice systems \cite{2009Santos}.}
\label{phase-diagram-BMF}
\end{figure}

Using the results of $\overline{a_{2}^{\left(0\right)}\left(i,0\right)}$ in Eq.(\ref{a2-avg}) and $\overline{a_{2}^{(2)}(i,0|i,0)}$ (see Appendix B), we obtain the beyond mean-field phase diagram of MI-BG in Fig. \ref{phase-diagram-BMF}.
For $d=2$ dimensions, from Fig. \ref{phase-diagram-BMF} (left) we can see that when disorder goes to zero ($\Delta/t=0$), our 2nd-order result (blue line) is exactly in agreement with that obtained from Landau effective potential theory in clean system (orange dot-dashed line)\cite{2009Santos}.
As disorder grows, higher disorder strength leads to smaller MI lobes, which is similar to the mean-field result.
Qualitatively similar behaviors can also be found in 3D, as shown in Fig. \ref{phase-diagram-BMF} (right).
The quantitative differences between 2D and 3D results hint that with the increase of the dimension $d$, the phase boundaries will approach the mean-field phase result. Such a behavior agrees with the observations in clean case \cite{2009Santos}, where mean field theory becomes exact in the limit $d\rightarrow \infty$ \cite{1999SSachdev}.

\begin{figure}[H]
\centering\includegraphics[width=8.2cm]{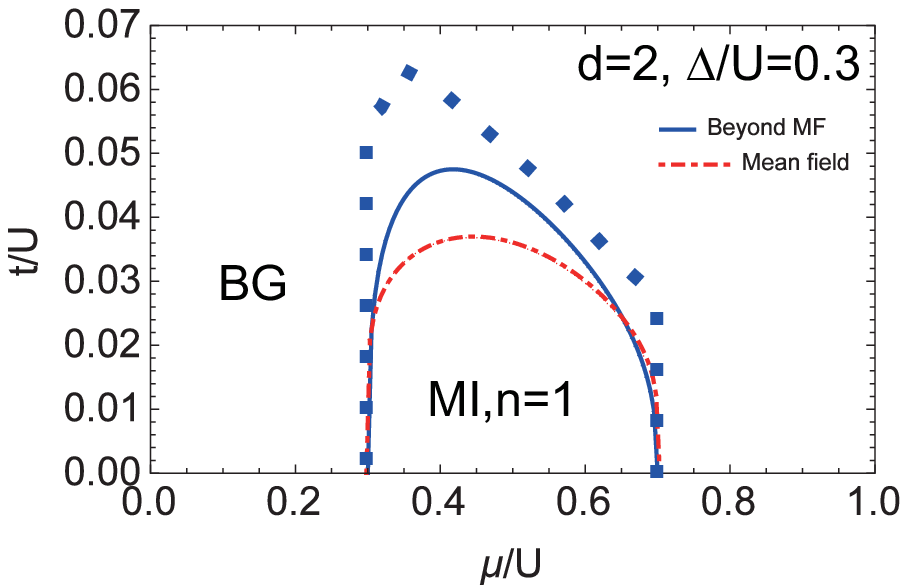}~~
\centering\includegraphics[width=7cm]{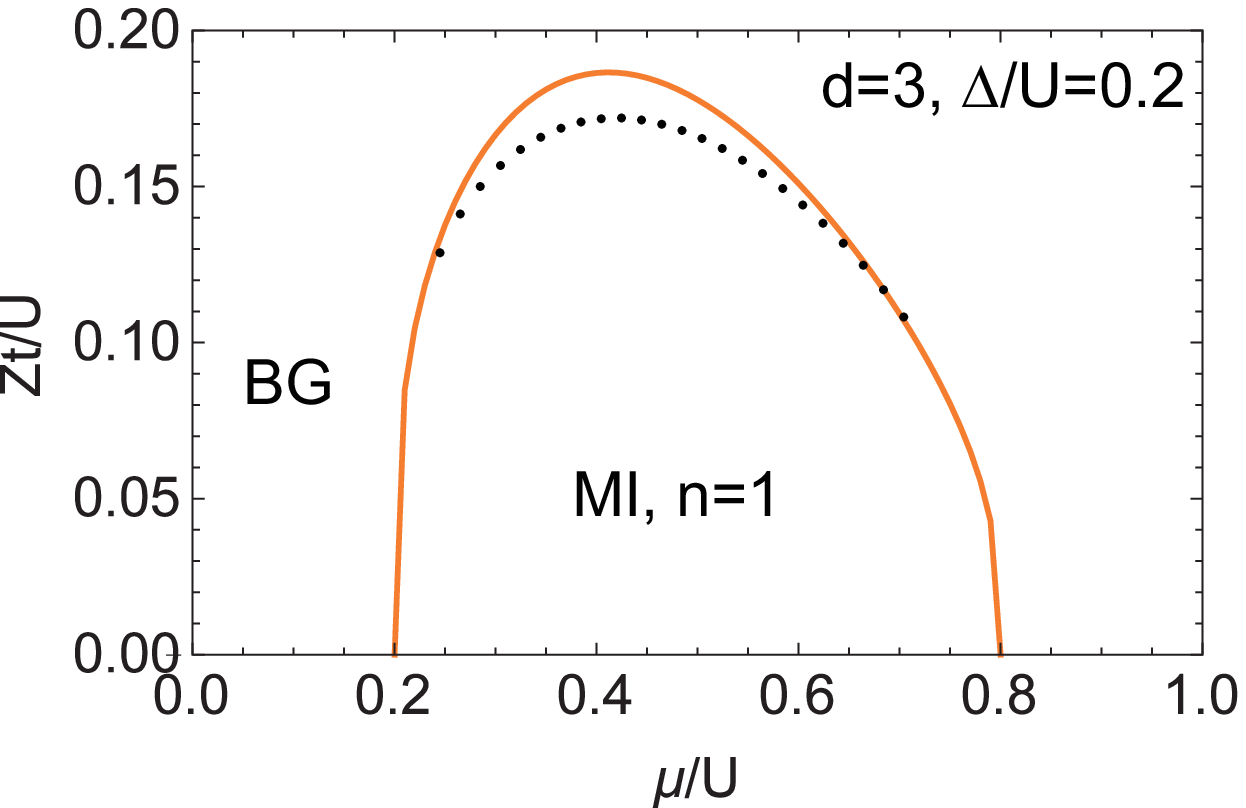}
\caption{(Color online) Comparison between our MI phase boundary and other data obtained from different methods with filling factor $n=1$. Left: MI-BG phase diagram in 2D with the disorder strength $\Delta/U=0.3$ at zero-temperature limit. Our beyond mean-field phase diagram is denoted by the blue line, and the red dot-dashed line corresponds to the mean-field result. The blue squares are the semianalytic results from canonical transformation approach \cite{2012CHLin}. Note that the disorder has been assumed to be uniformly distributed between $[-V/2,V/2]$, so their results at $V/U=0.6$ exactly correspond to that at $\Delta/U=0.3$ in our case. Right: MI-BG phase diagram in 3D with the disorder strength $\Delta/U=0.2$ at zero-temperature limit. The black dots coming from numerical integration within a renormalization group analysis \cite{2011Kruger}. The orange line denotes our analytic result beyond mean field approximation.}
\label{phase-diagram-comparison}
\end{figure}

In the following, we shall compare our results with the data found from other methods. More recently, based on a canonical transformation, a theoretical formalism has been developed to give the equilibrium phase diagram of clean BHM in a semianalytic way, which is a good match with the quantum Monte Carlo results  \cite{2012Projector}.
After some improvements, such a canonical transformation approach has been extended to the disordered case and yielded the phase diagram in 2D \cite{2012CHLin}. Their semianalytic results are denoted by the blue squares in Fig. \ref{phase-diagram-comparison} (left). Our mean-field result is represented by the red dot-dashed line, above which is our beyond mean-field result plotted by the blue line.
From this figure, a good tendency of approaching their semianalytic results can be observed.
Meanwhile, it can also be clearly seen that the beyond mean-field result (blue line) improves considerably the mean-field phase boundary (red dot-dashed line), indicating that the effect of quantum fluctuation indeed plays an important role in determining the phase boundary.
In  Fig. \ref{phase-diagram-comparison} (right), we give the comparison with our phase diagram of MI-BG with that obtained from a renormalization group analysis in 3D, where the relative variance of disorder-induced mass distribution is verified as the order parameter for MI-BG transition \cite{2011Kruger}. Their numerical integration result and our analytical beyond mean-field result are denoted by the black points and the orange line respectively. It is found that the deviation between our analytic calculation (orange line) and the data coming from numerical integration (black dots) is within 8.2 \%.

\section{Conclusion}

In this paper we have generalized the Ginzburg-Landau effective action approach to the disordered Bose-Hubbard Model.
Assuming that the vanishing of the modulus of parameter, i.e. $\left|\overline{\psi}\right|\equiv\left|\overline{\left\langle \hat{a}_{i}\right\rangle }\right|=0$ as the criteria to distinguish MI phase from BG phase, we have obtained the analytical beyond mean-field phase boundaries of MI-BG in 2D and 3D at zero temperature limit.
To the lowest order, our results coincide with that found from previous mean-field theory.
It turns out that the high-order correction improves considerably the phase boundary, and shows a good agreement with more recent semianalytical calculations.
The present approach could be generalized to include higher-order tunneling corrections, especially with the help of numerical simulation skills \cite{2009AEckardt,2009NTeichmann-process-chain}, so as to provide arbitrarily accurate results.
Meanwhile, it is also possible to take into account finite temperature effect within the present framework, which may contribute to the understanding of the competition between thermal fluctuations and quantum fluctuations.


\begin{acknowledgments}
We are grateful for stimulating discussions with Axel Pelster. The work is supported by National Natural Science Foundation of China (No. 11275119) and by Ph.D. Programs Foundation of Ministry of Education of China(No. 20123108110004). Support from Shanghai Key Laboratory of High Temperature Superconductors (No. 14DZ2260700) is also acknowledged.
\end{acknowledgments}

\appendix


\section{Fourth-order coefficient $a_{4}^{(0)}$}
Considering the time translation invariance of the cumulant $_{i}C_{4}^{(0)}\left(\tau_{1},\tau_{2}|\tau_{3},\tau_{4}\right)={}_{i}C_{4}^{(0)}\left(\tau_{1}-\tau_{3},\tau_{2}-\tau_{3}|0,\tau_{4}-\tau_{3}\right)$, together with the definition Eq.(\ref{a4_0-def}) and Matsubara transformation Eq.(\ref{matsubara-convention}), the fourth-order coefficient is expressed as
\begin{align}
  a_{4}^{(0)}(i,\omega_{m1};i,\omega_{m2}|i,\omega_{m4})=\frac{1}{\beta}\int_{0}^{\beta}d\tau_{1}\int_{0}^{\beta}d\tau_{2}\int_{0}^{\beta}d\tau_{4}{}_{i}C_{4}^{(0)}(\tau_{1},\tau_{2}|0,\tau_{4})e^{-i\left(\omega_{m1}\tau_{1}+\omega_{m2}\tau_{2}-\omega_{m4}\tau_{4}\right)},
\end{align}
where
\begin{align}
  _{i}C_{4}^{(0)}(\tau_{1},\tau_{2}|0,\tau_{4})= & \left<\hat{T}\left[\hat{a}_{i}^{\dagger}(\tau_{1})\hat{a}_{i}^{\dagger}(\tau_{2})\hat{a}_{i}(0)\hat{a}_{i}(\tau_{4})\right]\right>_{0}
  -_{i}C_{2}^{(0)}(\tau_{1}|0)_{i}C_{2}^{(0)}(\tau_{2}|\tau_{4})-_{i}C_{2}^{(0)}(\tau_{1}|\tau_{4})_{i}C_{2}^{(0)}(\tau_{2}|0).
\end{align}
The thermal average $\left<\hat{T}\left[\hat{a}_{i}^{\dagger}(\tau_{1})\hat{a}_{i}^{\dagger}(\tau_{2})\hat{a}_{i}(\tau_{4})\right]\hat{a}_{i}(0)\right>_{0}$ consists of the following 6 terms:
\begin{align}
  \vartheta\left(\tau_{1}-\tau_{2}\right)\vartheta\left(\tau_{2}-\tau_{4}\right)\left<\hat{a}_{i}^{\dagger}(\tau_{1})\hat{a}_{i}^{\dagger}(\tau_{2})\hat{a}_{i}(\tau_{4})\hat{a}_{i}(0)\right>_{0},~~\vartheta\left(\tau_{1}-\tau_{4}\right)\vartheta\left(\tau_{4}-\tau_{2}\right)\left<\hat{a}_{i}^{\dagger}(\tau_{1})\hat{a}_{i}(\tau_{4})\hat{a}_{i}^{\dagger}(\tau_{2})\hat{a}_{i}(0)\right>_{0}
  \nonumber \\
  \vartheta\left(\tau_{2}-\tau_{1}\right)\vartheta\left(\tau_{1}-\tau_{4}\right)\left<\hat{a}_{i}^{\dagger}(\tau_{2})\hat{a}_{i}^{\dagger}(\tau_{1})\hat{a}_{i}(\tau_{4})\hat{a}_{i}(0)\right>_{0},~~\vartheta\left(\tau_{2}-\tau_{4}\right)\vartheta\left(\tau_{4}-\tau_{1}\right)\left<\hat{a}_{i}^{\dagger}(\tau_{2})\hat{a}_{i}(\tau_{4})\hat{a}_{i}^{\dagger}(\tau_{1})\hat{a}_{i}(0)\right>_{0}
  \nonumber \\
  \vartheta\left(\tau_{4}-\tau_{1}\right)\vartheta\left(\tau_{1}-\tau_{2}\right)\left<\hat{a}_{i}(\tau_{4})\hat{a}_{i}^{\dagger}(\tau_{1})\hat{a}_{i}^{\dagger}(\tau_{2})\hat{a}_{i}(0)\right>_{0},~~\vartheta\left(\tau_{4}-\tau_{2}\right)\vartheta\left(\tau_{2}-\tau_{1}\right)\left<\hat{a}_{i}(\tau_{4})\hat{a}_{i}^{\dagger}(\tau_{2})\hat{a}_{i}^{\dagger}(\tau_{1})\hat{a}_{i}(0)\right>_{0}
  \label{thermal-term}
\end{align}

Taking the first term as example,
\begin{align}
  \left<\hat{a}_{i}^{\dagger}(\tau_{1})\hat{a}_{i}^{\dagger}(\tau_{2})\hat{a}_{i}(\tau_{4})\hat{a}_{i}(0)\right>_{0}=n\left(n-1\right)e^{\left[f_{i}\left(n\right)-f_{i}\left(n-1\right)\right]\tau_{1}}e^{\left[f_{i}\left(n-1\right)-f_{i}\left(n-2\right)\right]\tau_{2}}e^{\left[f_{i}\left(n-2\right)-f_{i}\left(n-1\right)\right]\tau_{4}},
\end{align}
we calculate
\begin{align}
  I_{124}= & \frac{1}{\beta} \int_{0}^{\beta}d\tau_{1}\int_{0}^{\beta}d\tau_{2}\int_{0}^{\beta}d\tau_{4}e^{-i\left(\omega_{m1}\tau_{1}+\omega_{m2}\tau_{2}-\omega_{m4}\tau_{4}\right)} \vartheta\left(\tau_{1}-\tau_{2}\right)\vartheta\left(\tau_{2}-\tau_{4}\right)\left<\hat{a}_{i}^{\dagger}(\tau_{1})\hat{a}_{i}^{\dagger}(\tau_{2})\hat{a}_{i}(\tau_{4})\hat{a}_{i}(0)\right>_{0}
  \nonumber \\ =&
  \frac{1}{\beta\mathcal{Z}_{i}^{(0)}}{\displaystyle \sum_{n=0}^{\infty}e^{-\beta f_{i}\left(n\right)}}\left\{ \delta_{\omega_{m_{2}},\omega_{m_{4}}}n\left(n-1\right)\left[\frac{\left(\beta e^{\beta\left(f_{i}\left(n\right)-f_{i}\left(n-1\right)-i\omega_{m_{1}}\right)}-\frac{e^{\beta\left(f_{i}\left(n\right)-f_{i}\left(n-1\right)-i\omega_{m_{1}}\right)}-1}{f_{i}\left(n\right)-f_{i}\left(n-1\right)-i\omega_{m_{1}}}\right)}{\left[f_{i}\left(n-2\right)-f_{i}\left(n-1\right)+i\omega_{m_{4}}\right]\left[f_{i}\left(n\right)-f_{i}\left(n-1\right)-i\omega_{m_{1}}\right]}\right.\right.
  \nonumber \\ &
  +\left.\frac{\left(\frac{e^{\beta\left(f_{i}\left(n\right)-f_{i}\left(n-2\right)-i\omega_{m_{4}}-i\omega_{m_{1}}\right)}-1}{f_{i}\left(n\right)-f_{i}\left(n-2\right)-i\omega_{m_{1}}-i\omega_{m_{4}}}-\frac{e^{\beta\left(f_{i}\left(n\right)-f_{i}\left(n-1\right)-i\omega_{m_{1}}\right)}-1}{f_{i}\left(n\right)-f_{i}\left(n-1\right)-i\omega_{m_{1}}}\right)}{\left[f_{i}\left(n-2\right)-f_{i}\left(n-1\right)+i\omega_{m_{4}}\right]^{2}}\right]
  \nonumber \\ &
  +\left(1-\delta_{\omega_{m_{2}},\omega_{m_{4}}}\right)n\left(n-1\right)\left[\frac{\left(\frac{e^{\beta\left(f_{i}\left(n\right)-f_{i}\left(n-1\right)-i\omega_{m_{3}}\right)}-1}{f_{i}\left(n\right)-f_{i}\left(n-1\right)-i\omega_{m_{3}}}-\frac{e^{\beta\left(f_{i}\left(n\right)-f_{i}\left(n-1\right)-i\omega_{m_{1}}\right)}-1}{f_{i}\left(n\right)-f_{i}\left(n-1\right)-i\omega_{m_{1}}}\right)}{\left[f_{i}\left(n-2\right)-f_{i}\left(n-1\right)+i\omega_{m_{4}}\right]\left(i\omega_{m_{4}}-\omega_{m_{2}}\right)}\right.
  \nonumber \\ &
  +\left.\left.\frac{\left(\frac{e^{\beta\left(f_{i}\left(n\right)-f_{i}\left(n-2\right)-i\omega_{m_{2}}-i\omega_{m_{1}}\right)}-1}{f_{i}\left(n\right)-f_{i}\left(n-2\right)-i\omega_{m_{1}}-i\omega_{m_{2}}}-\frac{e^{\beta\left(f_{i}\left(n\right)-f_{i}\left(n-1\right)-i\omega_{m_{1}}\right)}-1}{f_{i}\left(n\right)-f_{i}\left(n-1\right)-i\omega_{m_{1}}}\right)}{\left[f_{i}\left(n-2\right)-f_{i}\left(n-1\right)+i\omega_{m_{4}}\right]\left[f_{i}\left(n-2\right)-f_{i}\left(n-1\right)+i\omega_{m_{2}}\right]}\right]\right\} ,
\end{align}
where $\delta_{\omega_{m_{2}},\omega_{m_{4}}}$ is the Kronecker delta.
Remind that during the integration, we follow the complete integral form
\begin{align}
  \int_{0}^{\tau^{\prime}}d\tau e^{(a-b)\tau}=\delta_{a,b}\tau^{\prime}+(1-\delta_{a,b})\frac{e^{(a-b)\tau^{\prime}}-1}{a-b}
\end{align}
when $\tau^{\prime}$ becomes the subsequent integral variable.
After performing similar calculations for other terms in Eq.(\ref{thermal-term}), finally we give the following fourth-order coefficient

\begin{align}
  & a_{4}^{(0)}(i,\omega_{m1};i,\omega_{m2}|i,\omega_{m4})
  \nonumber \\  =&
  \delta_{\omega_{m_{2}},\omega_{m_{4}}}\frac{1}{\beta\mathcal{Z}_{i}^{(0)}}{\displaystyle \sum_{n=0}^{\infty}e^{-\beta f_{i}\left(n\right)}}
  \left\{ n\left(n-1\right)\left[\frac{\left(\beta e^{\beta\left(f_{i}\left(n\right)-f_{i}\left(n-1\right)-i\omega_{m_{1}}\right)}-\frac{e^{\beta\left(f_{i}\left(n\right)-f_{i}\left(n-1\right)-i\omega_{m_{1}}\right)}-1}{f_{i}\left(n\right)-f_{i}\left(n-1\right)-i\omega_{m_{1}}}\right)}{\left[f_{i}\left(n-2\right)-f_{i}\left(n-1\right)+i\omega_{m_{4}}\right]\left[f_{i}\left(n\right)-f_{i}\left(n-1\right)-i\omega_{m_{1}}\right]}\right.\right.
  \nonumber \\ &
  ~~~~~~~~~~~~~~~~~~~~~~~~~~+\left.\frac{\left(\frac{e^{\beta\left(f_{i}\left(n\right)-f_{i}\left(n-2\right)-i\omega_{m_{4}}-i\omega_{m_{1}}\right)}-1}{f_{i}\left(n\right)-f_{i}\left(n-2\right)-i\omega_{m_{1}}-i\omega_{m_{4}}}-\frac{e^{\beta\left(f_{i}\left(n\right)-f_{i}\left(n-1\right)-i\omega_{m_{1}}\right)}-1}{f_{i}\left(n\right)-f_{i}\left(n-1\right)-i\omega_{m_{1}}}\right)}{\left[f_{i}\left(n-2\right)-f_{i}\left(n-1\right)+i\omega_{m_{4}}\right]^{2}}\right]
  \nonumber \\ &
  ~~~~~~~~~~~~~~+n^{2}\left[\frac{\left(\beta e^{\beta\left(f_{i}\left(n\right)-f_{i}\left(n-1\right)-i\omega_{m_{1}}\right)}-\frac{e^{\beta\left(f_{i}\left(n\right)-f_{i}\left(n-1\right)-i\omega_{m_{1}}\right)}-1}{f_{i}\left(n\right)-f_{i}\left(n-1\right)-i\omega_{m_{1}}}\right)}{\left[f_{i}\left(n\right)-f_{i}\left(n-1\right)-i\omega_{m_{2}}\right]\left[f_{i}\left(n\right)-f_{i}\left(n-1\right)-i\omega_{m_{1}}\right]}\right.
  \nonumber \\ &
  ~~~~~~~~~~~~~~~~~~~~~+\left.\frac{\left(\frac{e^{i\beta\left(\omega_{m_{4}}-\omega_{m_{1}}\right)}-1}{i\omega_{m_{4}}-i\omega_{m_{1}}}-\frac{e^{\beta\left(f_{i}\left(n\right)-f_{i}\left(n-1\right)-i\omega_{m_{1}}\right)}-1}{f_{i}\left(n\right)-f_{i}\left(n-1\right)-i\omega_{m_{1}}}\right)}{\left[f_{i}\left(n\right)-f_{i}\left(n-1\right)-i\omega_{m_{2}}\right]^{2}}\right]
  \nonumber \\ &
  ~~~~~~~~
  +n\left(n+1\right)\left[\frac{\left(\frac{e^{\beta\left(f_{i}\left(n\right)-f_{i}\left(n-1\right)-i\omega_{m_{1}}\right)}-1}{f_{i}\left(n\right)-f_{i}\left(n-1\right)-i\omega_{m_{1}}}-\frac{e^{\beta\left(f_{i}\left(n\right)-f_{i}\left(n+1\right)+i\omega_{m_{4}}\right)}-1}{f_{i}\left(n\right)-f_{i}\left(n+1\right)+i\omega_{m_{4}}}\right)}{\left[f_{i}\left(n\right)-f_{i}\left(n-1\right)-i\omega_{m_{2}}\right]\left[f_{i}\left(n+1\right)-f_{i}\left(n-1\right)-i\omega_{m_{1}}-i\omega_{m_{2}}\right]}\right.
  \nonumber \\ &
  ~~~~~~~~~~~~~~~~~~~~~~~
  +\left.\left.\frac{\left(\frac{e^{i\beta\left(\omega_{m_{4}}-\omega_{m_{1}}\right)}-1}{i\omega_{m_{4}}-i\omega_{m_{1}}}-\frac{e^{\beta\left(f_{i}\left(n\right)-f_{i}\left(n+1\right)+i\omega_{m_{4}}\right)}-1}{f_{i}\left(n\right)-f_{i}\left(n+1\right)+i\omega_{m_{4}}}\right)}{\left[f_{i}\left(n\right)-f_{i}\left(n-1\right)-i\omega_{m_{2}}\right]\left[f_{i}\left(n\right)-f_{i}\left(n+1\right)+i\omega_{m_{1}}\right]}\right]\right\} _{\omega_{m1}\leftrightarrow\omega_{m2}}
  \nonumber \\  +&
  \left(1-\delta_{\omega_{m_{2}},\omega_{m_{4}}}\right)\frac{1}{\beta\mathcal{Z}_{i}^{(0)}}{\displaystyle \sum_{n=0}^{\infty}e^{-\beta f_{i}\left(n\right)}}
  \left\{ n\left(n-1\right)\left[\frac{\left(\frac{e^{\beta\left(f_{i}\left(n\right)-f_{i}\left(n-1\right)-i\omega_{m_{3}}\right)}-1}{f_{i}\left(n\right)-f_{i}\left(n-1\right)-i\omega_{m_{3}}}-\frac{e^{\beta\left(f_{i}\left(n\right)-f_{i}\left(n-1\right)-i\omega_{m_{1}}\right)}-1}{f_{i}\left(n\right)-f_{i}\left(n-1\right)-i\omega_{m_{1}}}\right)}{\left[f_{i}\left(n-2\right)-f_{i}\left(n-1\right)+i\omega_{m_{4}}\right]\left(i\omega_{m_{4}}-\omega_{m_{2}}\right)}\right.\right.
  \nonumber \\ &
  ~~~~~~~~~~~~~~~~~~~~~~~~~~~~~~~~~~+\left.\frac{\left(\frac{e^{\beta\left(f_{i}\left(n\right)-f_{i}\left(n-2\right)-i\omega_{m_{2}}-i\omega_{m_{1}}\right)}-1}{f_{i}\left(n\right)-f_{i}\left(n-2\right)-i\omega_{m_{1}}-i\omega_{m_{2}}}-\frac{e^{\beta\left(f_{i}\left(n\right)-f_{i}\left(n-1\right)-i\omega_{m_{1}}\right)}-1}{f_{i}\left(n\right)-f_{i}\left(n-1\right)-i\omega_{m_{1}}}\right)}{\left[f_{i}\left(n-2\right)-f_{i}\left(n-1\right)+i\omega_{m_{4}}\right]\left[f_{i}\left(n-2\right)-f_{i}\left(n-1\right)+i\omega_{m_{2}}\right]}\right]
  \nonumber \\ &
  ~~~~~~~~~~~~~~~~~~~~~+n^{2}\left[\frac{\left(\frac{e^{\beta\left(f_{i}\left(n\right)-f_{i}\left(n-1\right)-i\omega_{m_{3}}\right)}-1}{f_{i}\left(n\right)-f_{i}\left(n-1\right)-i\omega_{m_{3}}}-\frac{e^{\beta\left(f_{i}\left(n\right)-f_{i}\left(n-1\right)-i\omega_{m_{1}}\right)}-1}{f_{i}\left(n\right)-f_{i}\left(n-1\right)-i\omega_{m_{1}}}\right)}{\left[f_{i}\left(n\right)-f_{i}\left(n-1\right)-i\omega_{m_{2}}\right]\left[i\omega_{m_{4}}-i\omega_{m_{2}}\right]}\right.
  \nonumber \\ &
  ~~~~~~~~~~~~~~~~~~~~~~~~~~~~+\left.\frac{\left(\frac{e^{i\beta\left(\omega_{m_{4}}-\omega_{m_{1}}\right)}-1}{i\omega_{m_{4}}-i\omega_{m_{1}}}-\frac{e^{\beta\left(f_{i}\left(n\right)-f_{i}\left(n-1\right)-i\omega_{m_{1}}\right)}-1}{f_{i}\left(n\right)-f_{i}\left(n-1\right)-i\omega_{m_{1}}}\right)}{\left[f_{i}\left(n\right)-f_{i}\left(n-1\right)-i\omega_{m_{2}}\right]\left[f_{i}\left(n\right)-f_{i}\left(n-1\right)-i\omega_{m_{4}}\right]}\right]
  \nonumber \\ &
  ~~~~~~~~~~~~~~~~~~
  +n\left(n+1\right)\left[\frac{\left(\frac{e^{\beta\left(f_{i}\left(n\right)-f_{i}\left(n-1\right)-i\omega_{m_{3}}\right)}-1}{f_{i}\left(n\right)-f_{i}\left(n-1\right)-i\omega_{m_{3}}}-\frac{e^{\beta\left(f_{i}\left(n\right)-f_{i}\left(n+1\right)+i\omega_{m_{4}}\right)}-1}{f_{i}\left(n\right)-f_{i}\left(n+1\right)+i\omega_{m_{4}}}\right)}{\left[f_{i}\left(n\right)-f_{i}\left(n-1\right)-i\omega_{m_{2}}\right]\left[f_{i}\left(n+1\right)-f_{i}\left(n-1\right)-i\omega_{m_{1}}-i\omega_{m_{2}}\right]}\right.
  \nonumber \\ &
  ~~~~~~~~~~~~~~~~~~~~~~~~~~~~~~~~
  +\left.\left.\frac{\left(\frac{e^{i\beta\left(\omega_{m_{4}}-\omega_{m_{1}}\right)}-1}{i\omega_{m_{4}}-i\omega_{m_{1}}}-\frac{e^{\beta\left(f_{i}\left(n\right)-f_{i}\left(n+1\right)+i\omega_{m_{4}}\right)}-1}{f_{i}\left(n\right)-f_{i}\left(n+1\right)+i\omega_{m_{4}}}\right)}{\left[f_{i}\left(n\right)-f_{i}\left(n-1\right)-i\omega_{m_{2}}\right]\left[f_{i}\left(n\right)-f_{i}\left(n+1\right)+i\omega_{m_{1}}\right]}\right]\right\} _{\omega_{m1}\leftrightarrow\omega_{m2}}
  \nonumber \\ &
  -\left\{ a_{2}^{(0)}\left(i,\omega_{m1}|i,\omega_{m4}\right)a_{2}^{(0)}\left(i,\omega_{m2}\right)\right\} _{\omega_{m1}\leftrightarrow\omega_{m2}},
  \label{a4}
\end{align}
where $\{\bullet\}_{\omega_{m1}\leftrightarrow \omega_{m2}}$ denote a symmetrization in the variables $\omega_{m1}$ and $\omega_{m2}$.

  \section{Second-order correction coefficient $a_{2}^{(2)}(i,\omega_{m1}|i,\omega_{m2})$}
Similarly, with the help of the time translation invariance in the expression of cumulant $C_{4}^{(0)}$ in Eq.(\ref{c4_0}), 
together with the definition Eq.(\ref{a2-2}) and Matsubara transformation Eq.(\ref{matsubara-convention}), we express the second-order correction coefficient as
\begin{align}
  a_{2}^{(2)}(i,\omega_{1}|i,\omega_{4})= & \delta_{\omega_{1},\omega_{2}}a_{2}^{(2)}(i,\omega_{1}),
  \nonumber \\
  a_{2}^{(2)}(i,\omega_{m})= & \int_{0}^{\beta}d\tau_{1}\int_{0}^{\beta}d\tau_{2}\int_{0}^{\beta}d\tau_{3}e^{-i\omega_{m}\tau_{1}}{}_{i}C_{4}^{(0)}(\tau_{1},\tau_{3}|\tau_{2},0)_{j}C_{2}^{(0)}(\tau_{2}|\tau_{3})
 \end{align}

Inserting the expression of the cumulant $C_{2}^{(0)}$ in Eq. (\ref{c2_0}) and $C_{4}^{(0)}$ in Eqs. (\ref{c4_0}), (\ref{thermal-term}), a lengthy yet straightforward calculation give us the final expression of the second-order correction coefficient

\begin{align}
  & a_{2}^{(2)}(i,\omega_{m})
  \nonumber \\ = & \frac{1}{\mathcal{Z}_{i}^{(0)}\mathcal{Z}_{j}^{(0)}}\sum_{m=0,n=0}^{+\infty}e^{-\beta\left[f_{i}\left(n\right)+f_{j}\left(m\right)\right]}\left\{ \frac{\left(n+1\right)^{2}\left(m+1\right)}{\left[f_{i}\left(n+1\right)-f_{i}\left(n\right)+i\omega_{m}\right]^{2}\left[f_{j}\left(m+1\right)-f_{j}\left(m\right)+i\omega_{m}\right]}\right.
  \nonumber \\ &
  -\frac{\left(n+1\right)^{2}m}{\left[f_{i}\left(n+1\right)-f_{i}\left(n\right)-f_{j}\left(m\right)+f_{j}\left(m-1\right)\right]\left[f_{i}\left(n+1\right)-f_{i}\left(n\right)+i\omega_{m}\right]^{2}}
  \nonumber \\ &
  -\frac{\left(n+1\right)^{2}m}{\left[f_{i}\left(n+1\right)-f_{i}\left(n\right)-f_{j}\left(m\right)+f_{j}\left(m-1\right)\right]^{2}\left[f_{i}\left(n+1\right)-f_{i}\left(n\right)+i\omega_{m}\right]}
  \nonumber \\ &
  -\frac{\left(n+1\right)^{2}m}{\left[f_{i}\left(n\right)-f_{i}\left(n+1\right)-f_{j}\left(m-1\right)+f_{j}\left(m\right)\right]^{2}\left[f_{j}\left(m\right)-f_{j}\left(m-1\right)+i\omega_{m}\right]}
  \nonumber \\ &
  +\frac{\left(n+1\right)\left(n+2\right)m}{\left[f_{i}\left(n+1\right)-f_{i}\left(n\right)+i\omega_{m}\right]^{2}\left[f_{i}\left(n+2\right)-f_{i}\left(n\right)-f_{j}\left(m\right)+f_{j}\left(m-1\right)+i\omega_{m}\right]}
  \nonumber \\ &
  +\frac{\left(n+1\right)\left(n+2\right)m}{\left[f_{i}\left(n\right)-f_{i}\left(n+1\right)-f_{j}\left(m-1\right)+f_{j}\left(m\right)\right]^{2}\left[f_{i}\left(n+2\right)-f_{i}\left(n\right)-f_{j}\left(m\right)+f_{j}\left(m-1\right)+i\omega_{m}\right]}
  \nonumber \\ &
  -\frac{2\left(n+1\right)\left(n+2\right)m}{\left[f_{i}\left(n\right)-f_{i}\left(n+1\right)-f_{j}\left(m-1\right)+f_{j}\left(m\right)\right]\left[f_{i}\left(n+2\right)-f_{i}\left(n\right)-f_{j}\left(m\right)+f_{j}\left(m-1\right)+i\omega_{m}\right]\left[f_{i}\left(n+1\right)-f_{i}\left(n\right)+i\omega_{m}\right]}
  \nonumber \\ &
  +\frac{n^{2}\left(m+1\right)}{\left[f_{i}\left(n\right)-f_{i}\left(n-1\right)-f_{j}\left(m+1\right)+f_{j}\left(m\right)\right]^{2}\left[f_{j}\left(m+1\right)-f_{j}\left(m\right)+i\omega_{m}\right]}
  \nonumber \\ &
  -\frac{n\left(n+1\right)\left(m+1\right)}{\left[f_{i}\left(n-1\right)-f_{i}\left(n\right)-f_{j}\left(m\right)+f_{j}\left(m+1\right)\right]\left[f_{i}\left(n+1\right)-f_{i}\left(n\right)+i\omega_{m}\right]^{2}}
  \nonumber \\ &
  -\frac{n\left(n+1\right)\left(m+1\right)}{\left[f_{i}\left(n-1\right)-f_{i}\left(n\right)-f_{j}\left(m\right)+f_{j}\left(m+1\right)\right]^{2}\left[f_{i}\left(n+1\right)-f_{i}\left(n\right)+i\omega_{m}\right]}
  \nonumber \\ &
  -\frac{2n\left(n+1\right)\left(m+1\right)}{\left[f_{i}\left(n\right)-f_{i}\left(n-1\right)-f_{j}\left(m+1\right)+f_{j}\left(m\right)\right]\left[f_{j}\left(m+1\right)-f_{j}\left(m\right)+i\omega_{m}\right]\left[f_{i}\left(n+1\right)-f_{i}\left(n\right)+i\omega_{m}\right]}
  \nonumber \\ &
  -\frac{n^{2}\left(m+1\right)}{\left[f_{i}\left(n-1\right)-f_{i}\left(n\right)-f_{j}\left(m\right)+f_{j}\left(m+1\right)\right]\left[f_{i}\left(n\right)-f_{i}\left(n-1\right)+i\omega_{m}\right]^{2}}
  \nonumber \\ &
  +\frac{n^{2}\left(m+1\right)}{\left[f_{i}\left(n-1\right)-f_{i}\left(n\right)-f_{j}\left(m\right)+f_{j}\left(m+1\right)\right]^{2}\left[f_{i}\left(n\right)-f_{i}\left(n-1\right)+i\omega_{m}\right]}
  \nonumber \\ &
  -\frac{n^{2}m}{\left[f_{i}\left(n\right)-f_{i}\left(n-1\right)+i\omega_{m}\right]^{2}\left[f_{j}\left(m\right)-f_{j}\left(m-1\right)+i\omega_{m}\right]}
  \nonumber \\ &
  -\frac{2n\left(n+1\right)m}{\left[f_{i}\left(n\right)-f_{i}\left(n+1\right)+f_{j}\left(m\right)-f_{j}\left(m-1\right)\right]\left[f_{i}\left(n\right)-f_{i}\left(n-1\right)+i\omega_{m}\right]\left[f_{j}\left(m\right)-f_{j}\left(m-1\right)+i\omega_{m}\right]}
  \nonumber \\ &
  -\frac{n\left(n+1\right)m}{\left[f_{i}\left(n+1\right)-f_{i}\left(n\right)-f_{j}\left(m\right)+f_{j}\left(m-1\right)\right]\left[f_{i}\left(n\right)-f_{i}\left(n-1\right)+i\omega_{m}\right]^{2}}
  \nonumber
\end{align}
\begin{align}
   &
  +\frac{n\left(n+1\right)m}{\left[f_{i}\left(n+1\right)-f_{i}\left(n\right)-E_{k}+E_{k-1}\right]^{2}\left[f_{i}\left(n\right)-f_{i}\left(n-1\right)+i\omega_{m}\right]}
  \nonumber \\&
  -\frac{n\left(n-1\right)\left(m+1\right)}{\left[f_{i}\left(n\right)-f_{i}\left(n-1\right)+i\omega_{m}\right]^{2}\left[f_{i}\left(n\right)-f_{i}\left(n-2\right)+f_{j}\left(m\right)-f_{j}\left(m+1\right)+i\omega_{m}\right]}
  \nonumber \\ &
  -\frac{2n\left(n-1\right)\left(m+1\right)}{\left[f_{i}\left(n\right)-f_{i}\left(n-1\right)+f_{j}\left(m\right)-f_{j}\left(m+1\right)\right]\left[f_{i}\left(n\right)-f_{i}\left(n-1\right)+i\omega_{m}\right]\left[f_{i}\left(n\right)-f_{i}\left(n-2\right)+f_{j}\left(m\right)-f_{j}\left(m+1\right)+i\omega_{m}\right]}
  \nonumber \\ &
  \left.-\frac{n\left(n-1\right)\left(m+1\right)}{\left[f_{i}\left(n\right)-f_{i}\left(n-1\right)-f_{j}\left(m+1\right)+f_{j}\left(m\right)\right]^{2}\left[f_{i}\left(n\right)-f_{i}\left(n-2\right)-f_{j}\left(m+1\right)+f_{j}\left(m\right)+i\omega_{m}\right]}\right\} -\left[a_{2}^{\left(0\right)}\left(i,\omega_{m}\right)\right]^{3}
\end{align}

In the following, we need to perform disorder ensemble average of the above coefficient, i.e.
\begin{align}
  \overline{a_{2}^{(2)}(i,\omega_{m})}=\left(\frac{1}{2\Delta}\right)^{2}\int_{-\Delta}^{\Delta}\int_{-\Delta}^{\Delta}a_{2}^{(2)}(i,\omega_{m})d\delta\mu_{i}d\delta\mu_{j}.
\end{align}
Considering the complexity and length of the above expression, the disorder ensemble averaged coefficient $\overline{a_{2}^{(2)}(i,\omega_{m})}$ can be well approximated by means of numerical skills, e.g. doing the summation according to definition of integral.


\end{document}